\pdfoutput=1
\documentclass[11pt]{article}

\usepackage[preprint]{acl}

\usepackage{times}
\usepackage{latexsym}

\usepackage[T1]{fontenc}
\usepackage[utf8]{inputenc}

\usepackage{microtype}

\usepackage{inconsolata}

\usepackage{graphicx}

\usepackage{mathrsfs}
\usepackage{paralist}
\usepackage{booktabs}
\usepackage{multirow}
\usepackage{pgfplots}
\usepackage{scalefnt}
\usepackage{amsthm,amsmath,amssymb}
\usepackage{mathtools}
\usepackage{bm}
\usepackage{colortbl}
\usepackage{xcolor}
\usepackage{array}
\usepackage{stfloats}
\usepackage{tikz-dependency}
\usepackage{tikz}
\usepackage[edges]{forest}
\usepackage{siunitx}
\usepackage{graphicx}
\usepackage[ruled]{algorithm2e}
\usepackage{lineno}
\usepackage{subfigure}
\usepackage{helvet}
\usepackage{courier}
\usepackage{CJKutf8}
\usepackage{makecell}
\usepackage{nicematrix}
\usepackage{forest}
\usetikzlibrary{shapes.geometric}
\title{From Pre-trained Models to Large Language Models: A Comprehensive Survey of AI-Driven Psychological Computing}

\author{
  \textbf{Huiyao Chen\textsuperscript{1,2}} 
  \textbf{Ruimeng Liu\textsuperscript{1}} 
  \textbf{Yan Luo\textsuperscript{1}}
  \textbf{Jiawen Zhang\textsuperscript{3}}
  \\
  \textbf{Meishan Zhang\textsuperscript{1}}$\thanks{~~Corresponding author: Meishan Zhang.}$
  \textbf{Baotian Hu\textsuperscript{1,2}}
  \textbf{Min Zhang\textsuperscript{1,2}}
  \\
  \textsuperscript{1}Institute of Computing and Intelligence, Harbin Institute of Technology (Shenzhen), China
  \\
  \textsuperscript{2} Shenzhen Loop Area Institute (SLAI)
  \\
  \textsuperscript{3}The Hong Kong University of Science and Technology (Guangzhou), China
  \\
  \texttt{chenhy1018@gmail.com, littlepuppyq@gmail.com}
  \\
  \texttt{hitszly@gmail.com, mason.zms@gmail.com}
}

\pgfplotsset{compat=1.18} 
\begin{document}
\maketitle
%\tableofcontents
\begin{abstract}
The intersection of artificial intelligence and psychological science has experienced remarkable growth, with annual publications expanding from 859 papers in 2000 to 29,979 by 2025. However, this rapid evolution has created methodological fragmentation where similar computational techniques are independently developed across isolated psychological domains. This survey introduces the first systematic taxonomy that organizes AI-driven psychology tasks by computational processing patterns rather than application domains, categorizing them into four fundamental types: classification, regression, structured relational, and generative interactive tasks. Through analysis of over 300 representative works spanning the pre-trained model era and large language model era, we examine how computational approaches evolved from task-specific feature engineering to transfer learning and few-shot adaptation. We provide systematic coverage of datasets, evaluation metrics, and benchmarks while addressing fundamental challenges including interpretability, label uncertainty, privacy constraints, and cross-cultural validity. This computational perspective reveals transferable methodological patterns previously obscured by domain-centric organization, enabling systematic knowledge transfer and accelerated progress in computational psychology.
\end{abstract}

\section{Introduction}
\label{sec:introduction}

Artificial intelligence has fundamentally transformed how we study human psychology.
What began as modest efforts to automate psychological assessments has evolved into a comprehensive computational approach spanning mental health diagnosis, personality analysis, cognitive evaluation, and therapeutic intervention.
This transformation is evidenced by remarkable publication growth: the field expanded from 859 papers in 2000 to 29,979 papers by 2025, representing a 35-fold increase that reflects both the maturation of AI technologies and the increasing recognition that computational methods can address psychological questions at scales previously unattainable through conventional research paradigms.

\usepgfplotslibrary{groupplots}
\usetikzlibrary{patterns,backgrounds}
\definecolor{skyblue}{HTML}{6456FF}

\begin{figure}[t]
\centering
\resizebox{\columnwidth}{!}{
\begin{tikzpicture}
    \pgfkeys{/pgf/number format/.cd,fixed, fixed zerofill,precision=0}
    \begin{groupplot}[group style={group name=myplot,group size=1 by 1,horizontal sep=25pt,vertical sep=20pt,xlabels at=edge bottom, ylabels at=edge left},height=9cm,width=16cm]

    \nextgroupplot[
    ylabel={\textbf{Number of Publications ($\times 10^3$)}},
    ylabel style={font=\LARGE},
    xmin=1999, xmax=2026,
    ymin=0, ymax=35,
    xtick={2000,2010,2018,2023,2025},
    xticklabel style={font=\normalsize, rotate=45, /pgf/number format/precision=0, /pgf/number format/fixed},
    ytick={0,5,10,15,20,25,30},
    yticklabel style={font=\normalsize},
    ymajorgrids=true,
    xmajorgrids=true,
    grid style={dashed, gray!30},
    width=16cm,
    height=9cm,
    axis lines=left,
    tick align=outside,
    major tick length=3pt,
    minor tick length=1.5pt,
    every axis plot/.append style={
        mark=*,
        mark options={solid, scale=0.8}
    }
    ]

    % 阶段背景色
    \fill[skyblue!15, opacity=0.6] (axis cs:1999,0) rectangle (axis cs:2010,35);
    \fill[orange!8, opacity=0.6] (axis cs:2010,0) rectangle (axis cs:2018,35);
    \fill[red!8, opacity=0.6] (axis cs:2018,0) rectangle (axis cs:2026,35);

    % 主要数据线 (2000-2024)
    \addplot[
        color=black,
        mark=*,
        thick,
        line width=2pt
    ] coordinates {
        (2000,0.859) (2001,0.898) (2002,1.070) (2003,1.288) (2004,1.502) (2005,1.490) 
        (2006,1.789) (2007,2.149) (2008,2.616) (2009,2.941) (2010,2.687) (2011,2.666) 
        (2012,2.827) (2013,3.449) (2014,3.692) (2015,4.150) (2016,4.416) (2017,5.520) 
        (2018,7.556) (2019,9.986) (2020,11.572) (2021,13.843) (2022,17.966 ) (2023,18.613) 
        (2024,23.028) (2025,29.979)
    };
    
    % 阶段分隔线
    \addplot[gray, dashed, thick, forget plot] coordinates {(2010,0) (2010,35)};
    \addplot[gray, dashed, thick, forget plot] coordinates {(2018,0) (2018,35)};
    
    % 阶段标注 - 重新定位避免遮盖线条
    \node[align=center, font=\large, fill=skyblue!20, rounded corners=3pt, 
          inner sep=4pt] at (axis cs:2005,20) {
          \textcolor{skyblue}{\textbf{Phase I (2000-2010)}} \\
          \textcolor{black}{\normalsize Foundational Growth} \\
          \textcolor{black}{\normalsize 200-300 papers/year}
    };
    
    \node[align=center, font=\large, fill=orange!20, rounded corners=3pt,
          inner sep=4pt] at (axis cs:2014,15) {
          \textcolor{orange!80!black}{\textbf{Phase II (2010-2018)}} \\
          \textcolor{black}{\normalsize Accelerated Development} \\
          \textcolor{black}{\normalsize 500-600 papers/year}
    };
    
    \node[align=center, font=\large, fill=red!20, rounded corners=3pt,
          inner sep=4pt] at (axis cs:2022,4.2) {
          \textcolor{red!80!black}{\textbf{Phase III (2018-2025)}} \\
          \textcolor{black}{\normalsize Exponential Growth} \\
          \textcolor{black}{\normalsize 20\% annual growth}
    };
    
    % 关键技术发展标注
    % \node[pin=90:{\normalsize\textit{Deep Learning Era}}, pin distance=8mm,
    %       font=\tiny] at (axis cs:2013,4) {};
    
    % \node[pin=0:{\normalsize\textit{Transformer \& LLMs}}, pin distance=10mm,
    %       font=\tiny] at (axis cs:2019,11.5) {};
    
    % 特殊标记点
    \addplot[red, mark=*, mark size=3pt, only marks, forget plot] 
      coordinates {(2010,2.687) (2018,7.556) (2025,29.979)};
    
    % 添加数值标签到关键点
    \node[above, font=\large] at (axis cs:2000,0.859) {859};
    \node[above, font=\large] at (axis cs:2010,2.687) {2,687};
    \node[above, font=\large] at (axis cs:2017,7.556) {7,556};
    \node[above, font=\large] at (axis cs:2024,29.979) {29,979};
    
    \end{groupplot}
    
\end{tikzpicture}
}
\caption{Publication growth trajectory in AI-driven psychology research from 2000 to 2025, showing three distinct phases of evolution from foundational growth to exponential expansion.}
\label{fig:ai-psychology-evolution}
\end{figure}
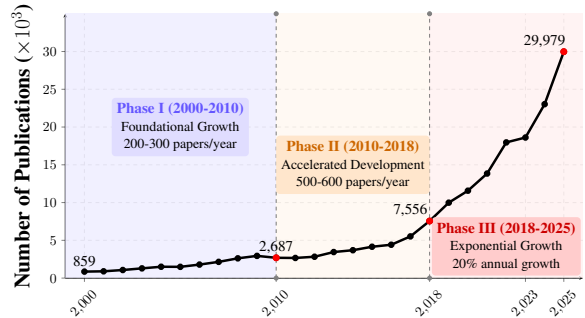

\usepgfplotslibrary{groupplots}
\usetikzlibrary{patterns,backgrounds}

% 使用推荐的十六进制颜色
\definecolor{lightblue}{HTML}{E2E0FF}    % 对应 blue!20
\definecolor{lightpink}{HTML}{FFE5CC}    % 对应 orange!20
\definecolor{darkred}{HTML}{FFCCCC}      % 对应 red!20

\begin{figure}[t]
\centering
\resizebox{\columnwidth}{!}{
\begin{tikzpicture}
    \pgfkeys{/pgf/number format/.cd,fixed, fixed zerofill,precision=0}
    \begin{axis}[
        width=22cm,
        height=12cm,
        ybar=2pt,            
        bar width=12pt,      
        % 使用对数刻度
        ymode=log,
        log basis y=10,
        ylabel={\textbf{Number of Publications (log scale)}},
        ylabel style={font=\LARGE},
        ymin=500,        
        ymax=100000,     
        % 对数刻度的刻度设置
        ytick={1000,5000,10000,50000,100000},
        yticklabels={1K,5K,10K,50K,100K},
        yticklabel style={font=\Large},
        xtick={1,2,3,4,5,6,7,8,9,10,11,12},
        xticklabels={
            {Computer\\Science},
            {Engineering},
            {Math \&\\Comp Bio},
            {Neurosciences\\Neurology},
            {Communication},
            {Mathematics},
            {Behavioral\\Sciences},
            {Psychology},
            {Radiology \&\\Medical Imaging},
            {Telecomm.},
            {Psychiatry},
            {Ophthalmology}
        },
        xticklabel style={font=\large, rotate=35, anchor=north east},
        x tick label style={align=center},
        xlabel={\textbf{Research Directions}},
        xlabel style={font=\LARGE},
        legend style={at={(0.98,0.98)}, anchor=north east, font=\Large, fill=white, fill opacity=0.9, text opacity=1},
        ymajorgrids=true,
        grid style={dashed, gray!30},
        enlarge x limits={abs=0.3},
        xmin=0.2,
        xmax=12.8,
        bar shift=0pt,
        axis lines=left,              % 添加这一行:只显示左边和下边的坐标轴
        tick align=outside            % 可选:让刻度线向外
    ]
    
    % 2000-2010 期间 - 浅蓝色填充，黑色边框
    \addplot[
        fill=lightblue,
        draw=lightblue,
        line width=1.2pt
    ] coordinates {
        (0.7,12776) (1.7,4460) (2.7,3981) (3.7,11749) (4.7,3318) (5.7,5672) 
        (6.7,12384) (7.7,12100) (8.7,3176) (9.7,806) (10.7,3542) (11.7,8152)
    };
    
    % 2010-2018 期间 - 浅粉色填充，黑色边框
    \addplot[
        fill=lightpink,
        draw=lightpink,
        line width=1.2pt
    ] coordinates {
        (1,23496) (2,11166) (3,9541) (4,15153) (5,8986) (6,9752) 
        (7,14060) (8,13106) (9,6505) (10,2983) (11,5094) (12,6440)
    };
    
    % 2018-2025 期间 - 暗红色填充，黑色边框
    \addplot[
        fill=darkred,
        draw=darkred,
        line width=1.2pt
    ] coordinates {
        (1.3,89188) (2.3,46075) (3.3,37840) (4.3,30548) (5.3,29638) (6.3,26037) 
        (7.3,24585) (8.3,23199) (9.3,18474) (10.3,13515) (11.3,13474) (12.3,3765)
    };
    
    \legend{2000-2010, 2010-2018, 2018-2025}
    
    \end{axis}
    
\end{tikzpicture}
}
\caption{Disciplinary shift in AI-driven psychology research across three developmental phases (2000-2025). The exponential growth in computer science publications (from $\sim$12K to $\sim$100K) compared to modest increases in traditional psychology domains illustrates the field's transformation from psychology-centric to computation-centric approaches.}
\label{fig:research-directions-evolution}
\end{figure}
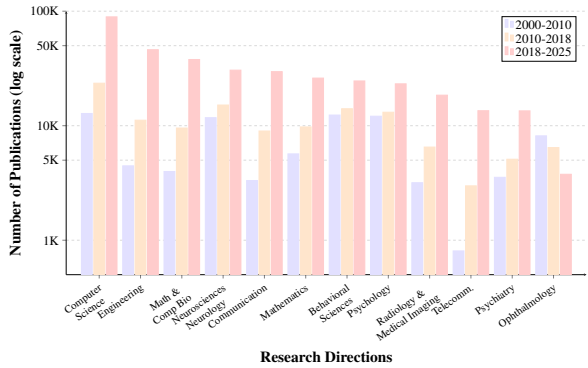

This growth trajectory reveals three distinct evolutionary phases, each characterized by different technological capabilities and research methodologies, as shown in Figure~\ref{fig:ai-psychology-evolution}.
The \textbf{foundational growth phase from 2000 to 2010} established basic computational frameworks for psychological data analysis through conventional machine learning approaches including support vector machines, decision trees, and ensemble methods.
Annual publications grew steadily from 859 to 2,687 papers, averaging 200 to 300 papers per year.
The \textbf{accelerated development phase from 2010 to 2018} witnessed the emergence of deep learning architectures, enabling more sophisticated pattern recognition through representation learning capabilities.
This period saw publications increase from 2,687 to 7,556 papers, with annual output reaching 500 to 600 papers per year.
The \textbf{exponential expansion phase from 2018 to 2025} began with the advent of pre-trained language models such as BERT and GPT-2, which initiated sustained annual growth rates exceeding 20 percent.
The introduction of large language models (LLMs) in 2023 further accelerated this trajectory, fundamentally transforming computational treatment of psychological tasks through zero-shot learning, few-shot adaptation, and natural language interfaces, ultimately reaching 29,979 papers by 2025.

This quantitative expansion reflects a fundamental disciplinary transformation in how psychological research is conducted.
Figure~\ref{fig:research-directions-evolution} reveals that computer science publications expanded exponentially from approximately 12,000 in the period spanning 2000 to 2010 to nearly 100,000 in the period spanning 2018 to 2025, representing an order of magnitude increase.
Engineering, mathematics, neurosciences, and medical imaging gained substantial prominence alongside conventional psychological disciplines.
In contrast, core psychology domains such as behavioral sciences and psychiatry showed modest linear growth.
This divergence indicates a fundamental shift from psychology-centric approaches toward \textbf{computation-centric methodologies} that leverage advanced AI techniques, multimodal data processing, and large-scale computational resources.
Understanding what makes psychological computing fundamentally different from conventional AI applications provides essential context for organizing this expanding field.

\begin{figure*}[t]
    \centering
    \includegraphics[width=1.0\linewidth]{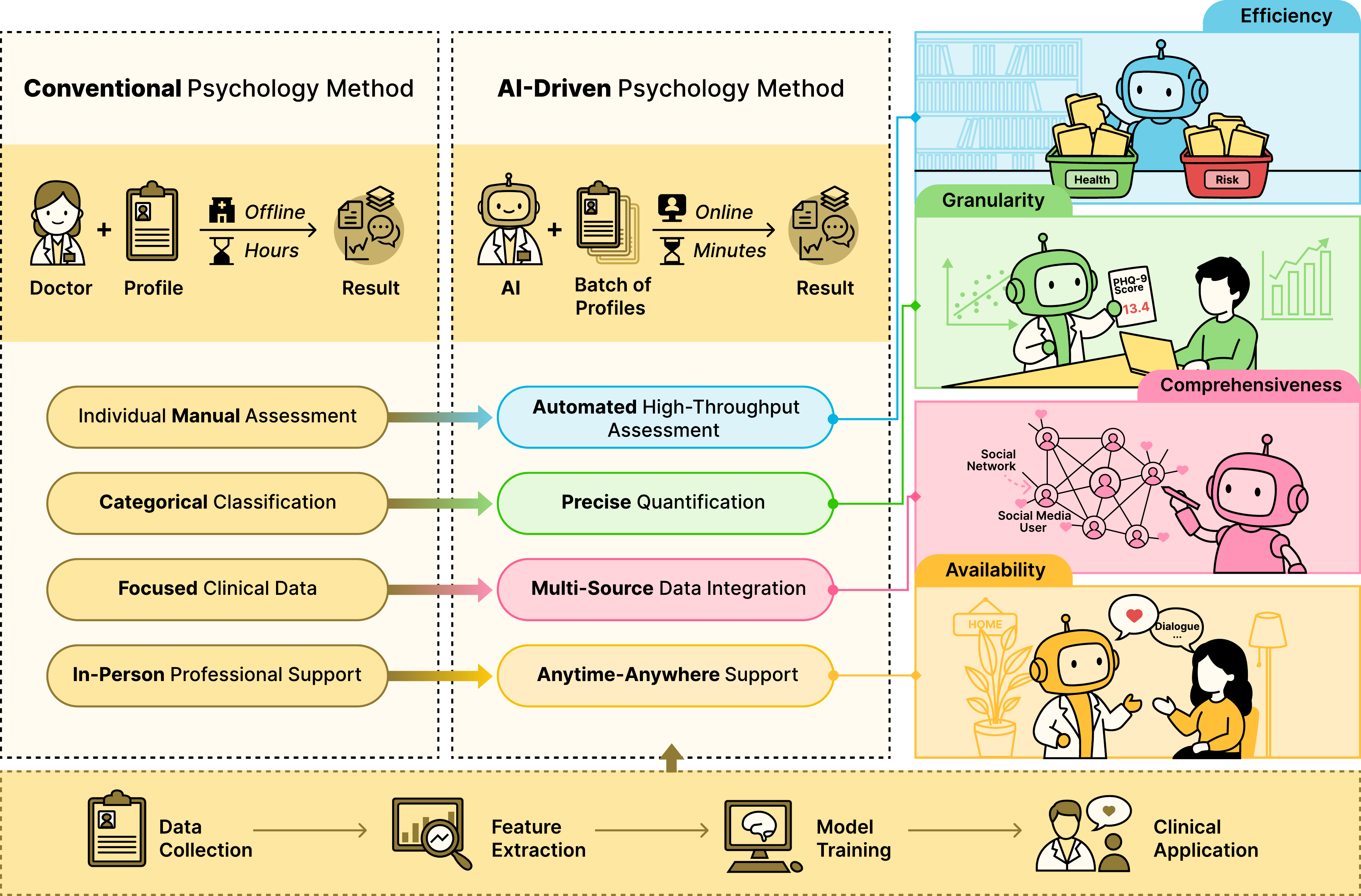}
    \caption{Evolution of psychology methods. Conventional approaches rely on individual manual assessment by trained professionals, providing focused clinical insights through in-person consultations. AI-driven methods complement these practices by enabling automated, high-throughput assessment, precise quantification of psychological states, multi-source data integration, and accessible, anytime-anywhere support, thereby enhancing the efficiency, granularity, comprehensiveness, and availability of psychological services.}
    \label{fig:framework}
\end{figure*}

Conventional psychological assessment methods have established rigorous clinical standards through decades of professional practice.
Trained clinicians conduct individual manual assessments, providing focused clinical insights through in-person consultations that capture nuanced behavioral patterns and contextual factors.
These conventional approaches excel at delivering personalized care and building therapeutic relationships essential for effective intervention.
However, the growing demand for psychological services and the emergence of diverse data sources have created opportunities for computational methods to complement these established practices.
As illustrated in Figure~\ref{fig:framework}, AI-driven approaches enhance conventional methods across four key dimensions.
First, \textbf{automated high-throughput assessment} enables simultaneous evaluation of large populations, processing batch profiles in minutes rather than hours, extending the reach of psychological services beyond the constraints of one-on-one consultations.
Second, \textbf{precise quantification} transforms categorical classifications into continuous measurements, such as converting binary depression screening into fine-grained severity scores like PHQ-9 assessments that track symptom progression over time.
Third, \textbf{multi-source data integration} synthesizes information from social networks, social media activity, and dialogue interactions alongside focused clinical data, providing a more comprehensive view of psychological states in naturalistic settings.
Fourth, \textbf{anytime-anywhere support} delivers accessible interventions through online platforms, complementing scheduled in-person professional consultations with continuous monitoring and immediate assistance during critical moments.

The successive waves of technological innovation driving this transformation have introduced progressively more powerful capabilities.
Early machine learning approaches demonstrated that statistical models could identify patterns in psychological data, moving beyond simple rule-based systems to capture complex relationships between behavioral indicators and mental states.
The introduction of pre-trained models like BERT and GPT-2 brought transfer learning capabilities, allowing researchers to leverage knowledge from large-scale datasets and adapt it to specialized psychological tasks with limited domain-specific data.
Most recently, LLMs such as GPT-4 and Claude have introduced capabilities for sophisticated psychological tasks with minimal task-specific training.
These models can now conduct structured clinical interviews, generate personalized therapeutic content, and analyze subtle linguistic markers of mental health conditions.

Yet this rapid expansion has created an increasingly fragmented research landscape.
Consider a researcher developing a transformer-based system for depression detection from social media posts.
They fine-tune BERT on labeled data, employ attention mechanisms to identify relevant linguistic features, and evaluate performance using clinical validation metrics.
Meanwhile, another team builds a personality assessment system using nearly identical technical approaches.
They use the same BERT architecture, similar fine-tuning strategies, and comparable attention-based feature extraction.
Despite these computational similarities, the two groups likely publish in different venues, use different terminologies, and remain largely unaware of each other's methodological innovations.

This pattern of methodological isolation repeats throughout the field.
Prompt engineering techniques that prove effective for therapeutic dialogue systems could benefit educational psychology applications, but such knowledge rarely transfers across domain boundaries.
Attention visualization methods developed for suicide risk prediction share fundamental mechanisms with approaches used in cognitive assessment, yet these methods are often rediscovered independently.
Researchers working on depression detection, personality assessment, and cognitive evaluation frequently employ similar neural architectures and computational techniques, yet cross-references between these subfields remain surprisingly rare.
The field has organized itself around application domains such as clinical psychology, educational psychology, and personality research rather than computational characteristics.
This organization obscures the technical commonalities that could accelerate progress.

This fragmentation carries tangible consequences for research efficiency and scientific progress.
Methodological advances remain confined to narrow application areas when they could benefit the broader field.
Researchers waste effort rediscovering techniques already established in adjacent domains.
Perhaps most critically, the lack of a unified computational framework makes it difficult to identify which technical approaches work well across different psychological tasks and which are fundamentally limited to specific applications.
Without a systematic way to organize and compare computational methods, the field risks continued balkanization as new AI technologies emerge.

\subsection{A Computational Perspective on Psychological Tasks}

This survey addresses these challenges by organizing AI-driven psychology through computational characteristics rather than application domains.
We propose that psychological AI applications can be understood through four fundamental computational paradigms.
Each paradigm is defined by how it processes information and generates outputs.

\textbf{Classification tasks} identify discrete psychological states through discriminative pattern recognition.
These tasks determine whether someone exhibits depression, identify personality types, or categorize cognitive states.
The computational challenge lies in learning decision boundaries that separate psychological constructs in high-dimensional feature spaces.
A depression detection classifier and a personality type identifier may serve different clinical purposes, but they face identical computational challenges.
These challenges include managing class imbalance, ensuring robust feature extraction, and maintaining performance across demographic groups.

\textbf{Regression tasks} quantify continuous psychological measurements by predicting numerical values along graded scales.
Rather than assigning categories, these tasks estimate depression severity scores, predict cognitive ability levels, or measure emotional intensity.
The computational focus shifts to learning functions that accurately map inputs to continuous output spaces while maintaining clinical interpretability.
Systems that predict depression severity and those that assess cognitive decline share the same technical requirements.
These requirements include handling measurement uncertainty and ensuring predictions align with established clinical scales.

\textbf{Structured relational tasks} extract and model complex relationships between psychological constructs through structured tuples.
These tasks extract entity-relation pairs, event-argument structures, or temporal sequences from unstructured data.
Outputs take the form of tuples that can construct symptom networks, social support graphs, or developmental timelines.
The computational challenge involves learning representations that capture relational structures and temporal dynamics while reflecting psychological theory.
Whether extracting symptom interactions from clinical narratives or modeling developmental trajectories in educational settings, these systems employ similar graph-based architectures and information extraction techniques.

\textbf{Generative interactive tasks} create personalized psychological content through dynamic engagement with users.
This category encompasses systems that explain psychological mechanisms, generate therapeutic interventions, provide mental health support, or deliver adaptive educational instruction.
We treat these as a unified computational paradigm because they share a fundamental mechanism: synthesizing contextually appropriate responses adapted to evolving user states.
Whether generating a single therapeutic recommendation or maintaining a multi-turn counseling dialogue, these systems perform the same core operation of contextual response synthesis.
A therapeutic dialogue system and an educational tutoring agent may target different applications, but both must solve the same technical problems.
These problems include maintaining context across interactions, generating appropriate responses, and adapting to user feedback.

This computational taxonomy reveals patterns invisible from a domain-centric perspective.
By organizing tasks around computational commonalities rather than application domains, we enable systematic knowledge transfer across conventionally isolated research communities.
Attention mechanisms developed for one task type can be adapted to others sharing the same computational structure.
Training strategies that address data scarcity in clinical applications can be transferred to educational contexts facing similar constraints.
Evaluation frameworks designed for one domain can be adapted to assess systems in related domains with analogous requirements.

\subsection{Survey Scope and Contributions}

Our analysis spans two transformative technological eras that have reshaped psychological research methodologies.
The \textbf{pre-trained model era spanning 2018 to 2022} introduced transfer learning through architectures like BERT and GPT-2, enabling sophisticated analysis with limited domain-specific data.
The \textbf{large language model era beginning in 2023} brought capabilities that dramatically reduced deployment barriers for psychological applications through zero-shot learning and natural language interfaces.
We examine how computational approaches evolved across these eras, tracing the progression from task-specific feature engineering to transfer learning paradigms and finally to adaptation capabilities that require minimal domain-specific training.
This historical perspective illuminates not only what changed but why certain approaches succeeded while others faced fundamental limitations.

We also address the growing importance of multimodal approaches that integrate text, speech, visual, and physiological signals.
Human psychological states manifest simultaneously across multiple communicative channels, and contemporary assessment systems increasingly leverage this multidimensional nature of human expression.
Our analysis examines how advances in computer vision, speech processing, and sensor technologies complement language-based models to create more comprehensive psychological assessment systems.

This survey makes four primary contributions to the computational psychology literature:

\begin{compactitem}

\item \textbf{Computational Taxonomy.} We introduce a comprehensive taxonomy organizing AI-driven psychology tasks by fundamental computational processing patterns rather than application domains in Section~\ref{sec:taxonomy}.
This taxonomy reveals transferable methodological patterns and enables systematic knowledge transfer across conventionally isolated subfields.

\item \textbf{Era-Spanning Technical Analysis.} Through analysis of over 300 representative works, we trace how computational approaches evolved from pre-trained models to LLMs in Section~\ref{sec:methods}.
We examine both single-modal and multimodal approaches, addressing how integration of text, speech, visual, and physiological signals creates more comprehensive assessment systems.

\item \textbf{Systematic Resource Survey.} We provide comprehensive coverage of datasets, evaluation metrics, and benchmarks across different psychological domains in Section~\ref{sec:datasets}.
Our survey helps researchers identify appropriate resources and understand tradeoffs between different evaluation approaches.

\item \textbf{Challenge Analysis and Future Directions.} We systematically analyze fundamental challenges in psychological computing including interpretability, privacy preservation, cultural validity, and clinical utility in Section~\ref{sec:challenges}.
For each challenge, we examine concrete technical approaches and promising research directions in Section~\ref{sec:future}.

\end{compactitem}

The remainder of this survey is organized as follows.
Section~\ref{sec:background} establishes theoretical and technical foundations, examining the unique characteristics of psychological data and formal task formulation.
Section~\ref{sec:taxonomy} presents our computational taxonomy with comprehensive analysis of representative applications and technical requirements for each task category.
Section~\ref{sec:datasets} surveys available datasets and evaluation metrics across different psychological domains.
Section~\ref{sec:methods} provides detailed technical analysis of methods spanning both technological eras, examining architectures, training strategies, and adaptation techniques.
Section~\ref{sec:challenges} identifies fundamental technical challenges and solution strategies.
Section~\ref{sec:future} discusses promising research directions and emerging opportunities.
Section~\ref{sec:conclusion} synthesizes key insights and provides recommendations for researchers and practitioners.

We maintain a public repository and consistently update related works and resources on GitHub\footnote{\href{https://github.com/DreamH1gh/Awesome-AI4Psychological-Papers}{github.com/DreamH1gh/Awesome-AI4Psychological-Papers}}.

\section{Background}
\label{sec:background}

This section establishes the theoretical and technical foundations for our computational taxonomy.
We first examine what makes psychological computing different from conventional AI applications (Section~\ref{sec:psych_foundations}), 
then present a formal task formulation that unifies the diverse landscape of psychological tasks (Section~\ref{sec:formal_formulation}).

\subsection{Characteristics of Psychological Computing}
\label{sec:unique_characteristics}

Psychological computing is not simply another AI application domain.
It carries constraints rooted in the nature of human psychology, the ethics of working with human subjects, and the complexity of mental phenomena.
Understanding these constraints is essential before introducing any computational framework.
We organize them into three layers: the psychological principles that define the problem space (Section~\ref{sec:psych_foundations}), the data properties that emerge from those principles (Section~\ref{sec:data_characteristics}), and the technical requirements that follow (Section~\ref{sec:distinctions}).

\subsubsection{Psychological Foundations}
\label{sec:psych_foundations}

The technical challenges in psychological computing are rooted in fundamental principles of psychological science established through decades of theoretical and empirical research.
Understanding these foundations is essential for appreciating why certain computational problems in psychology are inherently complex rather than merely engineering obstacles.

\textbf{Construct validity} means that a model must measure what it claims to measure.
Psychological constructs such as depression or anxiety are theoretical entities that cannot be directly observed.
\citet{allport1937personality} formalized this as the tension between general laws that apply across populations and the unique patterns that characterize each individual.
\citet{cronbach1955construct} established that any measurement must be validated against the construct it targets, not merely against a proxy label.
A model predicting depression from sleep patterns may be capturing fatigue rather than the core affective symptoms of the disorder.
\citet{cronbach1957two} later noted that individual differences in psychology are meaningful signals, not noise to be averaged away.
Computational systems must therefore operate at both levels of analysis.

\textbf{Ecological validity} means that the same behavior can carry different psychological meanings depending on context and culture.
\citet{bronfenbrenner1977toward} showed that human behavior is shaped by multiple layers of context, from immediate social settings to broader cultural systems.
Cross-cultural research~\cite{triandis1989self} confirms that psychological constructs manifest differently across populations, which challenges the universality assumptions built into most machine learning systems.
A model trained on one population cannot be assumed to generalize to another without explicit validation.
\citet{brunswik2023perception} argued that models must reflect the real environments in which behavior occurs.
A person who withdraws from social contact may be experiencing depression, processing grief, or following cultural norms around solitude.

\textbf{Measurement uncertainty} is inherent to psychological assessment and cannot be eliminated through better annotation.
\citet{meehl1954clinical} showed that trained clinicians regularly disagree on diagnoses for the same patient, not because of error but because of legitimate differences in interpretation.
Self-report data introduces further distortions, as individuals may underreport stigmatized behaviors or lack accurate introspective access to their own states~\cite{crowne1960new, nisbett1977telling}.
The phenomenological tradition~\cite{giorgi2009descriptive} holds that mental states are experienced and interpreted differently by each individual, making objective ground truth difficult to define.
Label disagreement in psychological datasets therefore reflects genuine ambiguity in the underlying phenomena, not noise to be cleaned away.

\textbf{Ethical and relational constraints} shape what data can be collected and how systems must behave.
The Belmont Report~\cite{united1978belmont} and the APA Ethics \citet{code2017ethical} require informed consent, confidentiality, and careful consideration of potential harms.
Research on therapeutic alliance~\cite{horvath2011alliance} shows that effective psychological interventions depend on an evolving relationship between practitioner and client.
\citet{norcross2011psychotherapy} describes this as therapeutic flexibility, meaning that interventions must adjust in response to how the client is progressing.
For computational systems, this means that static deployment models are insufficient and that systems must remain transparent enough to preserve human oversight and clinical trust.

Together, these four principles explain why psychological computing presents challenges that go beyond standard engineering problems.

\subsubsection{Data Characteristics}
\label{sec:data_characteristics}

The psychological principles above translate into four concrete properties of psychological data that create distinctive challenges for model design.

\textbf{Multi-scale temporal dynamics.}
A person's mood can shift within minutes, while personality traits remain stable over decades.
Anxiety may spike during a stressful event and return to baseline within hours, yet the underlying tendency persists across many situations.
A model must capture both short-term fluctuations and long-term stability, while recognizing that the same behavioral signal can mean different things at different times.
A student spending more time alone may be depressed, deeply focused on coursework, or following culturally normal patterns during exam season.
This multi-scale, context-sensitive nature is more demanding than the single-scale temporal modeling typical in speech recognition or the time-independent classification common in image analysis.

\textbf{Systematic label uncertainty.}
When multiple clinicians assess the same patient, they may reach different conclusions based on their training and interpretation of ambiguous symptoms.
Self-report instruments introduce additional distortions: a patient may underreport substance use, overestimate medication adherence, or sincerely believe they are functioning well when clinical indicators suggest otherwise.
Cultural variation adds another layer of complexity.
Depression in Western clinical settings is typically characterized by sadness and emotional withdrawal, while in some East Asian contexts it more commonly presents through physical complaints such as fatigue and pain.
Standard annotation pipelines that resolve disagreement through majority voting discard this meaningful variation.
Models that treat labels as ground truth without accounting for annotator disagreement will learn a distorted picture of the underlying constructs.

\textbf{Privacy and data scarcity.}
Psychological data includes some of the most sensitive personal information that exists, covering mental health diagnoses, trauma histories, and suicidal ideation.
Collecting and sharing such data requires ethical review, informed consent, and compliance with regulations such as HIPAA and GDPR.
In practice, this limits dataset sizes, prevents free sharing across institutions, and requires technical safeguards such as de-identification and differential privacy.
These constraints have no close parallel in computer vision or general natural language processing, where large-scale training data can often be collected from public sources with minimal restriction.

\textbf{Meaningful individual variation.}
Population-level patterns in psychological data frequently fail to generalize to specific individuals or subgroups.
A depression detection model trained on college students may perform poorly on elderly populations, who express distress differently and face distinct life circumstances.
A personality model validated on Western samples may misclassify individuals from collectivist cultures where self-presentation norms differ substantially.
Effective systems must explicitly model individual-level variation rather than treating it as deviation from a population mean.

\subsubsection{Technical Distinctions}
\label{sec:distinctions}

The characteristics described above lead to three concrete departures from standard machine learning practice.

\textbf{Multi-dimensional evaluation.}
Standard machine learning relies on a single aggregate metric such as accuracy or F$_1$ score.
Psychological systems require evaluation across multiple dimensions simultaneously.
A depression detection model that achieves 95\% accuracy on a held-out test set may still fail in important ways.
It may capture sleep disruption and fatigue rather than the core affective features of depression, meaning it lacks construct validity.
It may produce accurate predictions without offering any guidance for treatment decisions, meaning it lacks clinical utility.
It may perform well on structured clinical interviews but fail on naturalistic social media text, meaning it lacks ecological validity.
It may perform well on average while performing poorly for specific demographic groups, raising fairness concerns.
Because objective ground truth is often unavailable, evaluation must incorporate expert judgment, longitudinal validation, and theoretical coherence alongside held-out test performance.

\textbf{Theory-aligned interpretability.}
In most AI applications, interpretability is a desirable property that aids debugging or builds user trust.
In psychological computing, it serves two more fundamental purposes.
Scientifically, interpretable models can generate falsifiable hypotheses that contribute to psychological theory.
Clinically, practitioners must be able to explain decisions to patients in terms that connect to established clinical concepts.
Telling a patient that a model flagged high activation in a particular neural network layer has no clinical meaning.
Explaining that the model identified patterns consistent with rumination and reduced engagement in previously enjoyed activities connects directly to diagnostic criteria and supports a meaningful clinical conversation.
This requirement often favors models with built-in interpretability over models that maximize predictive performance at the cost of transparency.

\begin{table*}[ht]
\fontsize{9}{10}\selectfont
\setlength{\tabcolsep}{1.3mm}

\begin{center}
\begin{NiceTabular*}{\textwidth}{@{\extracolsep{\fill}}lccc}[
    code-before = \rowcolor{blue!15}{1} \rowcolor{gray!15}{3}
]
\Xhline{0.08em}
\multicolumn{4}{c}{\textit{\textbf{Multi-dimensional Evaluation}}} \\

\bf Aspect & \bf Traditional AI & \bf Psychological Computing & \bf Why It Matters \\
\hline
Performance Metrics & Accuracy, F$_1$-score & \makecell[c]{Construct validity,\\ clinical utility,\\ ecological validity, fairness} & \makecell[c]{High accuracy $\neq$ \\ measuring the right construct} \\
Ground Truth & Objective labels & \makecell[c]{Expert judgment with \\ legitimate disagreement} & \makecell[c]{Label disagreement reflects \\ genuine ambiguity, not noise; \\ single metrics are insufficient} \\
\end{NiceTabular*}

\begin{NiceTabular*}{\textwidth}{@{\extracolsep{\fill}}lccc}[
    code-before = \rowcolor{blue!15}{1} \rowcolor{gray!15}{3}
]
\specialrule{.1em}{.05em}{0.05em}
\multicolumn{4}{c}{\textit{\textbf{Theory-Aligned Interpretability}}} \\

\bf Aspect & \bf Traditional AI & \bf Psychological Computing & \bf Why It Matters \\
\hline
Explainability & \makecell[c]{Post-hoc for \\ high-stakes cases} & \makecell[c]{Theory-aligned for \\ scientific and clinical use} & \makecell[c]{Models must generate \\ falsifiable hypotheses and \\ support clinical conversations} \\
Feature Alignment & Statistical patterns & \makecell[c]{Map to psychological \\ constructs} & \makecell[c]{Explanations must connect \\ to clinical concepts (e.g., anhedonia), \\ not statistical artifacts} \\
\end{NiceTabular*}

\begin{NiceTabular*}{\textwidth}{@{\extracolsep{\fill}}lccc}[
    code-before = \rowcolor{blue!15}{1} \rowcolor{gray!15}{3}
]
\specialrule{.1em}{.05em}{0.05em}
\multicolumn{4}{c}{\textit{\textbf{Continuous and Personalized Adaptation}}} \\

\bf Aspect & \bf Traditional AI & \bf Psychological Computing & \bf Why It Matters \\
\hline
Model Lifecycle & Static after training & \makecell[c]{Continuous learning \\ across timescales} & \makecell[c]{Symptom profiles change \\ during treatment; constructs \\ shift across the lifespan} \\
Personalization & Population-level patterns & \makecell[c]{Individual-level modeling \\ across subgroups} & \makecell[c]{Population patterns fail \\ for individuals; same behavior \\ carries different meanings \\ across contexts and cultures} \\
\bottomrule
\end{NiceTabular*}
\caption{
Key distinctions between traditional AI and psychological computing across three technical dimensions.
}
\label{tab:ai_psychology_distinctions}
\end{center}
\end{table*}

\textbf{Continuous adaptation.}
Many AI systems are trained once and deployed without further modification.
Psychological computing systems must support continuous adaptation.
A patient's symptom profile changes during treatment.
Coping strategies learned in therapy alter the behavioral patterns a model was trained to recognize.
Psychological constructs manifest differently across the lifespan, from childhood through adolescence to old age.
A system that cannot update its representations in response to these changes will become progressively less useful over time.
This requires online learning mechanisms that can incorporate new individual-specific information without degrading performance on previously learned patterns.

Table~\ref{tab:ai_psychology_distinctions} summarizes these distinctions across three dimensions, comparing psychological computing with conventional AI applications in terms of evaluation criteria, interpretability requirements, and adaptation demands.

These characteristics collectively explain why psychological computing cannot be addressed by applying standard machine learning pipelines to psychological data.

\subsection{Formal Task Formulation}
\label{sec:formal_formulation}

\begin{figure*}[t]
    \centering
    \includegraphics[width=\linewidth]{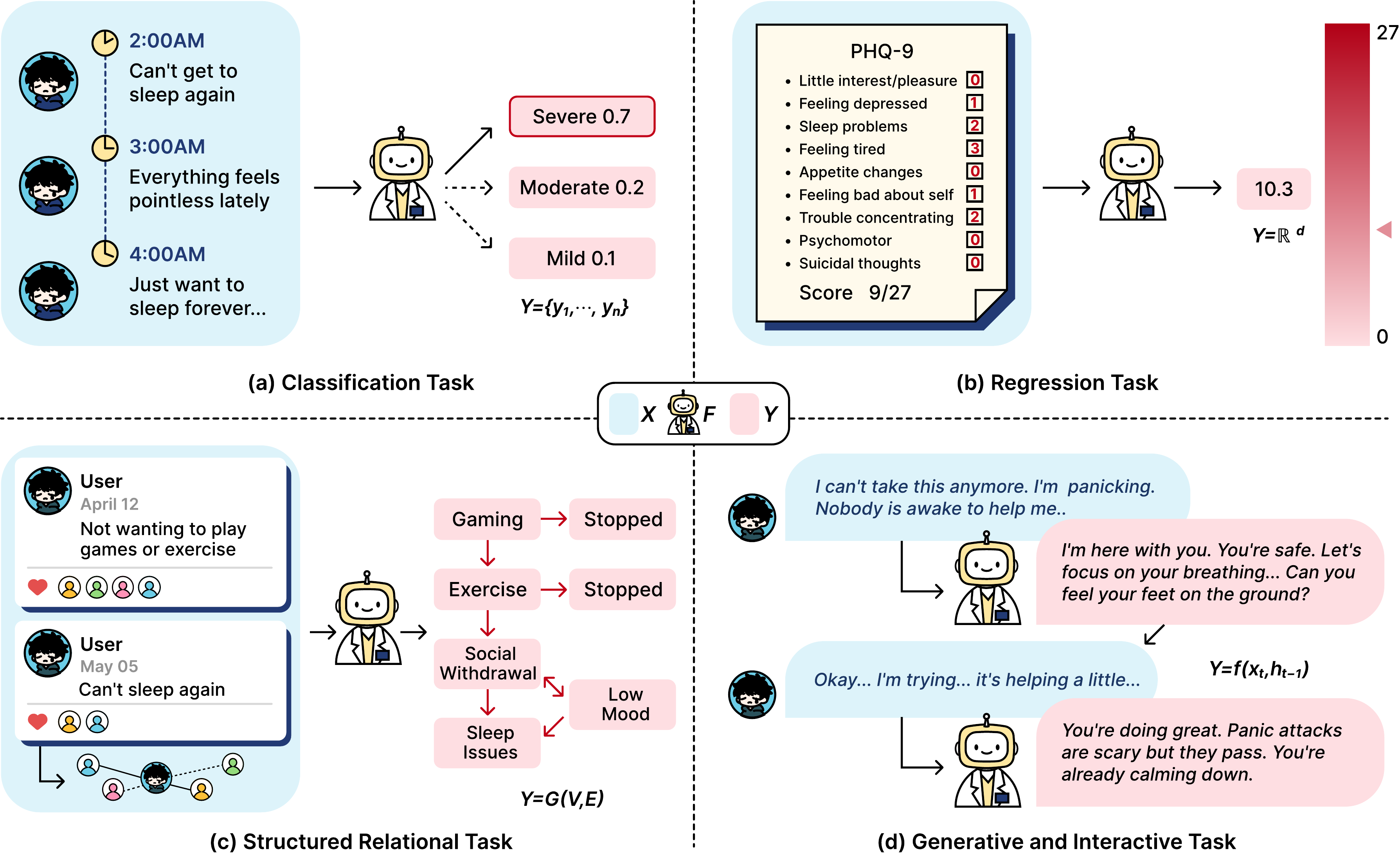}
    \caption{AI-driven psychology task computational framework. The figure illustrates four main task categories with their corresponding input modalities, network architectures, and output types: (a) Classification tasks for discrete psychological state identification, (b) Regression tasks for continuous psychological measurement, (c) Structured relational tasks for extracting psychological entities and relations as structured tuples, and (d) Generative interactive tasks for dynamic content creation and user engagement.
}
    \label{fig:task_formulate}
\end{figure*}

Building on the unique characteristics of psychological computing established in Section~\ref{sec:unique_characteristics}, we now present a unified mathematical framework that formalizes the computational characteristics of psychological computing tasks. This framework addresses the psychological foundations in Section~\ref{sec:psych_foundations}, data characteristics in Section~\ref{sec:data_characteristics}, and technical distinctions in Section~\ref{sec:distinctions}. It provides the foundation for our task-oriented taxonomy in Section~\ref{sec:taxonomy} by explicitly capturing how different psychological objectives translate into distinct computational requirements.

AI-driven psychology tasks can be formally characterized as a tuple $\mathcal{T} = \langle \mathcal{X}, \mathcal{Y}, \mathcal{F} \rangle$, where:

\begin{compactitem}
    \item \textbf{Input space} $\mathcal{X}$ encompasses diverse modalities of human-generated data. Text from clinical interviews or social media posts provides linguistic markers of psychological states through word choice, sentiment, and discourse patterns. Behavioral signals captured through smartphone sensors or wearable devices reveal activity patterns, sleep-wake cycles, and social interactions. Physiological measurements including heart rate variability and electrodermal activity reflect autonomic nervous system responses to stress and emotion. Multimedia content such as facial expressions in video or prosodic features in speech convey emotional states through nonverbal channels. Contextual information about time, location, and social setting provides essential interpretive frameworks for understanding whether behaviors are normative or clinically significant. For tasks requiring historical context or multi-turn interactions, the input space includes temporal sequences that capture how these signals evolve over time.
    
    \item \textbf{Output space} $\mathcal{Y}$ represents theoretically grounded psychological constructs rather than arbitrary labels. As illustrated in Figure~\ref{fig:task_formulate}, four distinct output types correspond to different computational requirements. 
    \begin{compactitem}
    \item Classification outputs take the form $\mathcal{Y} = \{y_1, \ldots, y_K\}$ for discrete categorical assignments. Figure~\ref{fig:task_formulate}(a) shows how late-night social media posts expressing sleep difficulties and feelings of pointlessness are processed to identify depression severity levels. These tasks address diagnostic categorization, personality assessment, and behavioral pattern identification.
    
    \item Regression outputs use $\mathcal{Y} = \mathbb{R}^d$ to represent continuous measurements. Figure~\ref{fig:task_formulate}(b) demonstrates how structured questionnaire responses from PHQ-9 are transformed into continuous severity scores. These tasks quantify symptom severity, stress levels, and cognitive abilities along graded scales.
    
    \item Structured relational outputs employ $\mathcal{Y} = \mathcal{G}(\mathcal{V}, \mathcal{E})$ to capture complex relationships between psychological constructs through structured tuples. These tasks extract entity-relation pairs, event-argument structures, or temporal sequences from unstructured data, producing outputs such as (entity$_1$, relation, entity$_2$) or (event, role, argument). Figure~\ref{fig:task_formulate}(c) illustrates how clinical narratives are processed to extract structured information that constructs symptom networks, social support graphs, or developmental timelines. These tasks reveal how psychological constructs interact and evolve through explicit relational structures.
    
    \item Generative interactive outputs produce dynamic sequences $\mathcal{Y}_t = f(x_t, h_{t-1})$ that create personalized content while adapting based on user engagement. Figure~\ref{fig:task_formulate}(d) shows how crisis messages receive immediate grounding responses, followed by adaptive therapeutic guidance that evolves based on user feedback. This unified category recognizes that both single-turn content generation and multi-turn interactive engagement fundamentally represent iterative content generation conditioned on evolving conversational states.
    \end{compactitem}
    
    \item \textbf{Computational function} $\mathcal{F}: \mathcal{X} \rightarrow \mathcal{Y}$ maps inputs to outputs while capturing complex relationships between observable data and psychological states. To address individual variability, this function incorporates personalization mechanisms. Individual-specific parameters allow the function to adapt decision boundaries or regression coefficients based on personal history and characteristics. Meta-learning approaches enable rapid adaptation to new individuals by learning how to learn from limited individual-specific data. For generative interactive tasks, $\mathcal{F}$ includes memory mechanisms that maintain representations of dialogue history and user preferences, along with state management systems that track evolving user needs and therapeutic goals across multiple interactions.
\end{compactitem}

This formalization encompasses the fundamental computational characteristics that distinguish psychological computing from traditional AI applications. The diversity in output types reflects the multifaceted nature of psychological phenomena, ranging from simple categorical classifications to complex multi-turn dialogues. The need to process multimodal inputs addresses the reality that psychological states manifest simultaneously across linguistic, behavioral, physiological, and social channels. The requirement to handle temporal dependencies captures how psychological processes unfold over time. The incorporation of contextual factors acknowledges that psychological expression depends on individual and situational contexts. The emphasis on theoretically grounded outputs ensures construct validity rather than mere predictive accuracy. Together, these requirements create computational challenges that necessitate specialized approaches beyond standard machine learning paradigms.

All task categories share common underlying principles. Multimodal data processing integrates diverse information sources to capture the multifaceted nature of psychological expression. Context-aware modeling accounts for individual and situational factors that modulate psychological manifestations. Output generation aligns with psychological theory and clinical practice rather than optimizing purely for predictive accuracy.

This formal framework establishes the conceptual foundation for organizing the diverse landscape of psychological computing applications. Section~\ref{sec:taxonomy} builds upon this formulation to present a comprehensive taxonomy that categorizes tasks according to their fundamental computational processing patterns rather than superficial application domains. By organizing tasks based on their output types, we reveal transferable methodological patterns and enable systematic knowledge transfer across traditionally isolated psychological subfields. For each task category, we detail the specific model architectures, training paradigms, and evaluation methodologies required, while explicating the computational-psychological mapping that makes these tasks tractable for AI systems while respecting the unique characteristics outlined in Section~\ref{sec:unique_characteristics}.

\section{A Computational Taxonomy}
\label{sec:taxonomy}

\subsection{Taxonomy Design Principles}

\tikzstyle{my-box}=[
    rectangle,
    draw=gray,
    rounded corners,
    text opacity=1,
    minimum height=1.5em,
    minimum width=5em,
    inner sep=2pt,
    align=center,
    fill opacity=.5,
    line width=0.8pt,
]
\tikzstyle{leaf}=[my-box, minimum height=1.5em,
    fill=pink!10, text=black, align=left,font=\LARGE,
    inner xsep=2pt,
    inner ysep=4pt,
    line width=0.8pt,
]

\definecolor{c0}{HTML}{7ac7e2} % color of root‘line
\definecolor{c1}{HTML}{e3716e} % Classification
\definecolor{c2}{HTML}{eca680} % Regression
\definecolor{c3}{HTML}{f7df87} % Structured Relational
\definecolor{c4}{HTML}{54beaa} % Generative and Interactive
\definecolor{c5}{HTML}{7ac7e2} % root

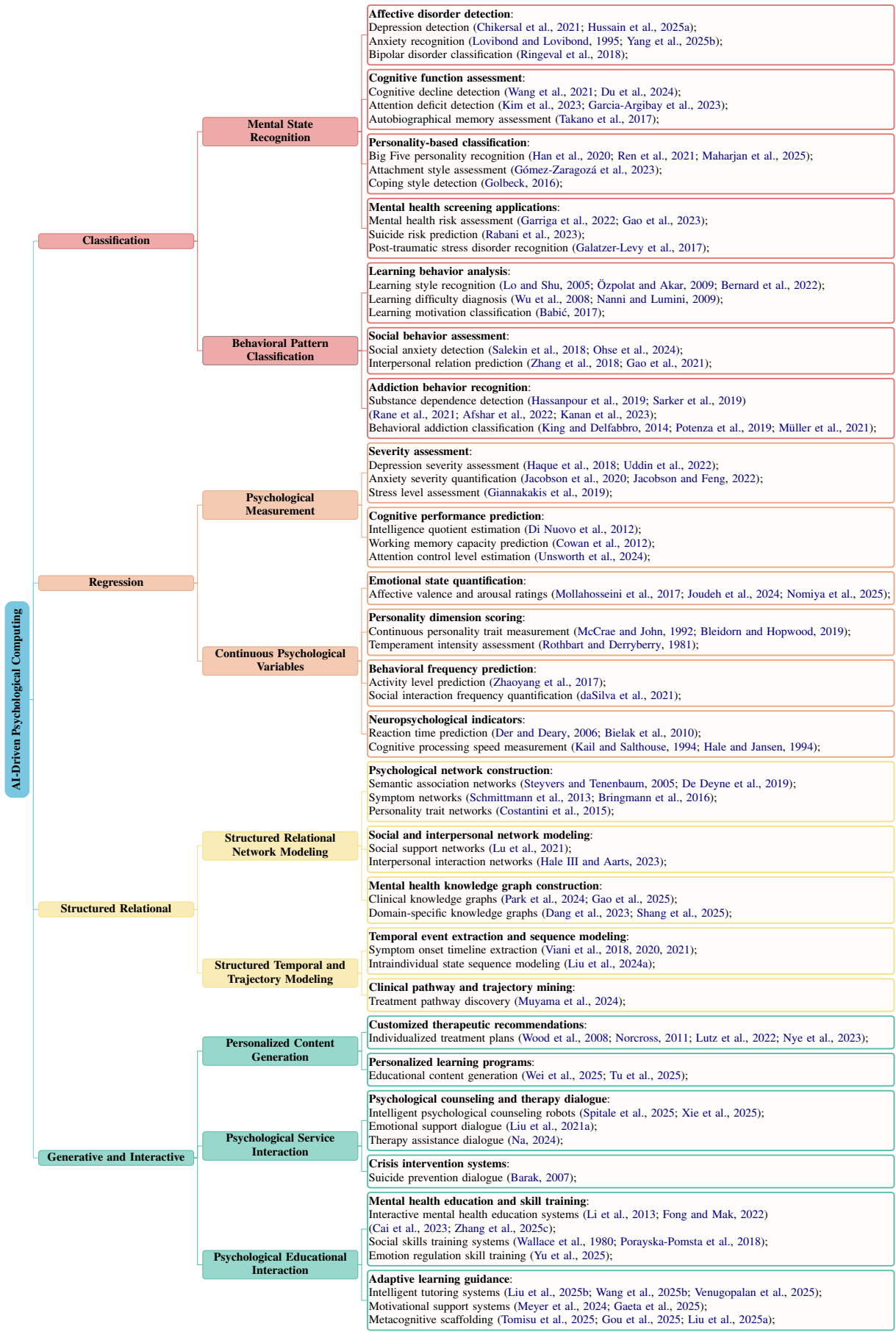
\begin{figure*}[!htp]
    \centering
    \resizebox{1\linewidth}{!}{
        \begin{forest}
            for tree={
                grow=east,
                reversed=true,
                anchor=west,
                parent anchor=east,
                child anchor=west,
                font=\LARGE,
                rectangle,
                draw=gray,
                rounded corners,
                align=left,
                text centered,
                minimum width=4em,
                s sep=5pt,
                inner xsep=4pt,
                inner ysep=1pt,
                line width=0.8pt,
                ver/.style={rotate=90, child anchor=north, parent anchor=south, anchor=center},
                calign=center, 
            },
            forked edges,
            where level=1{text width=20em,font=\LARGE,}{},
            where level=2{text width=20
            em,font=\LARGE,align=center}{},
            where level=3{text width=70em,font=\LARGE,}{},
            [
                \textbf{AI-Driven Psychological Computing}, ver, fill=c5, draw=c5, line width=5mm
                [
                    \textbf{Classification}, fill=c1!60, draw=c1, line width=0mm, edge={c0}
                    [
                        \textbf{Mental State}\\ \textbf{Recognition}, fill=c1!60, draw=c1, line width=0mm, edge={c1}
                        [ 
                            \textbf{Affective disorder detection}:\\ Depression detection \cite{chikersal2021detecting,hussain2025multi};\\
                            Anxiety recognition \cite{lovibond1995structure,yang2025concise};\\
                            Bipolar disorder classification \cite{ringeval2018avec};
                            , leaf, draw=c1, line width=0.7mm, edge={c1}
                        ]
                        [
                            \textbf{Cognitive function assessment}:\\
                            Cognitive decline detection \cite{wang2021development,du2024enhancing};\\
                            Attention deficit detection \cite{kim2023machine,garcia2023predicting};\\
                            Autobiographical memory assessment \cite{takano2017unraveling};
                            , leaf, draw=c1, line width=0.7mm, edge={c1}
                        ]
                        [
                            \textbf{Personality-based classification}:\\
                            Big Five personality recognition \cite{han2020knowledge,ren2021sentiment,maharjan2025psychometric};\\
                            Attachment style assessment  \cite{gomez2023online};\\
                            Coping style detection  \cite{golbeck2016detecting};
                            , leaf, draw=c1, line width=0.7mm, edge={c1}
                        ]
                        [
                            \textbf{Mental health screening applications}:\\
                            Mental health risk assessment \cite{garriga2022machine,gao2023machine};\\
                            Suicide risk prediction \cite{rabani2023detecting};\\
                            Post-traumatic stress disorder recognition \cite{galatzer2017utilization};\\
                            , leaf, draw=c1, line width=0.7mm, edge={c1}
                        ]
                    ]
                    [
                        \textbf{Behavioral Pattern} \\ 
                        \textbf{Classification}, fill=c1!60, line width=0mm, edge={c1}
                        [
                            \textbf{Learning behavior analysis}:\\
                            Learning style recognition \cite{lo2005identification,ozpolat2009automatic,bernard2022improving};\\
                            Learning difficulty diagnosis \cite{wu2008evaluation,nanni2009ensemble};\\
                            Learning motivation classification \cite{babic2017machine};
                            , leaf, draw=c1, line width=0.7mm, edge={c1}
                        ]
                        [
                            \textbf{Social behavior assessment}:\\
                            Social anxiety detection \cite{salekin2018weakly,ohse2024gpt};\\
                            Interpersonal relation prediction \cite{zhang2018facial,gao2021multi};
                            , leaf, draw=c1, line width=0.7mm, edge={c1}
                        ]
                        [
                            \textbf{Addiction behavior recognition}:\\
                            Substance dependence detection \cite{hassanpour2019identifying,sarker2019machine}\\ \cite{rane2021aim,afshar2022development,kanan2023intelligent};\\
                            Behavioral addiction classification \cite{king2014cognitive,potenza2019gambling,muller2021object};
                            , leaf, draw=c1, line width=0.7mm, edge={c1}
                        ]
                    ]
                ]
                [
                    \textbf{Regression}, fill=c2!60, draw=c2, line width=0mm, edge={c0}
                    [
                        \textbf{Psychological}\\ \textbf{Measurement}, fill=c2!60, draw=c2, line width=0mm, edge={c2}
                        [
                            \textbf{Severity assessment}:\\
                            Depression severity assessment \cite{haque2018measuring,uddin2022deep};\\
                            Anxiety severity quantification \cite{jacobson2020digital,jacobson2022anxiety};\\
                            Stress level assessment \cite{giannakakis2019review};
                            , leaf, draw=c2, line width=0.7mm, edge={c2}
                        ]
                        [
                            \textbf{Cognitive performance prediction}:\\
                            Intelligence quotient estimation \cite{di2012intelligent};\\
                            Working memory capacity prediction \cite{cowan2012models};\\
                            Attention control level estimation \cite{unsworth2024individual};
                            , leaf, draw=c2, line width=0.7mm, edge={c2}
                        ]
                    ]
                    [
                        \textbf{Continuous Psychological}\\
                        \textbf{Variables},fill=c2!60, draw=c2, line width=0mm, edge={c2}
                        [
                            \textbf{Emotional state quantification}: \\
                            Affective valence and arousal ratings
                            \cite{mollahosseini2017affectnet,joudeh2024predicting,nomiya2025artificial};
                            , leaf, draw=c2, line width=0.7mm, edge={c2}
                        ]
                        [
                            \textbf{Personality dimension scoring}:\\
                            Continuous personality trait measurement \cite{mccrae1992introduction,bleidorn2019using};\\
                            Temperament intensity assessment \cite{rothbart1981development};
                            , leaf, draw=c2, line width=0.7mm, edge={c2}
                        ]
                        [
                            \textbf{Behavioral frequency prediction}:\\
                            Activity level prediction \cite{zhaoyang2017morning};\\
                            Social interaction frequency quantification \cite{dasilva2021daily};
                            , leaf, draw=c2, line width=0.7mm, edge={c2}
                        ]
                        [
                            \textbf{Neuropsychological indicators}:\\
                            Reaction time prediction \cite{der2006age,bielak2010intraindividual};\\
                            Cognitive processing speed measurement \cite{kail1994processing,hale1994global};
                            , leaf, draw=c2, line width=0.7mm, edge={c2}
                        ]
                    ]
                ]
                [
                    \textbf{Structured Relational}, fill=c3!60, draw=c3, line width=0mm, edge={c0}
                    [
                        \textbf{Structured Relational}\\ \textbf{Network Modeling},fill=c3!60, draw=c3, line width=0mm, edge={c3}
                        [
                            \textbf{Psychological network construction}:\\
                            Semantic association networks \cite{steyvers2005large,de2019small};\\
                            Symptom networks \cite{schmittmann2013deconstructing,bringmann2016assessing};\\
                            Personality trait networks \cite{costantini2015state};
                            , leaf, draw=c3, line width=0.7mm, edge={c3}
                        ]
                        [
                            \textbf{Social and interpersonal network modeling}:\\
                            Social support networks \cite{lu2021development};\\
                            Interpersonal interaction networks \cite{hale2023hidden};                          
                            , leaf, draw=c3, line width=0.7mm, edge={c3}
                        ]
                        [
                            \textbf{Mental health knowledge graph construction}:\\
                            Clinical knowledge graphs \cite{park2024leveraging,gao2025large};\\
                            Domain-specific knowledge graphs \cite{dang2023gena,shang2025integrating};
                            , leaf, draw=c3, line width=0.7mm, edge={c3}
                        ]
                    ]
                    [
                        \textbf{Structured Temporal and}\\ \textbf{Trajectory Modeling},fill=c3!60, draw=c3, line width=0mm, edge={c3}
                        [
                            \textbf{Temporal event extraction and sequence modeling}:\\
                            Symptom onset timeline extraction \cite{viani2018time,viani2020temporal,viani2021natural};\\
                            Intraindividual state sequence modeling  \cite{liu2024intraindividual};
                            , leaf, draw=c3, line width=0.7mm, edge={c3}
                        ]
                        [
                            \textbf{Clinical pathway and trajectory mining}:\\
                            Treatment pathway discovery \cite{muyama2024machine};
                            , leaf, draw=c3, line width=0.7mm, edge={c3}
                        ]
                    ]
                ]
                [
                    \textbf{Generative and Interactive}, fill=c4!60, draw=c4, line width=0mm, edge={c0}
                    [
                        \textbf{Personalized Content}\\ \textbf{Generation}, fill=c4!60, draw=c4, line width=0mm, edge={c4}
                        [
                            \textbf{Customized therapeutic recommendations}:\\
                            Individualized treatment plans \cite{wood2008adapting,norcross2011psychotherapy,lutz2022prospective,nye2023efficacy};
                            , leaf, draw=c4, line width=0.7mm, edge={c4}
                        ]
                        [
                            \textbf{Personalized learning programs}:\\
                            Educational content generation \cite{wei2025advancement,tu2025empowering};
                            , leaf, draw=c4, line width=0.7mm, edge={c4}
                        ]
                    ]
                    [
                        \textbf{Psychological Service}\\ \textbf{Interaction}, fill=c4!60, draw=c4, line width=0mm, edge={c4}
                        [
                            \textbf{Psychological counseling and therapy dialogue}:\\
                            Intelligent psychological counseling robots \cite{spitale2025vita,xie2025psydt};\\
                            Emotional support dialogue \cite{liu-etal-2021-towards};\\
                            Therapy assistance dialogue \cite{na-2024-cbt};
                            , leaf, draw=c4, line width=0.7mm, edge={c4}
                        ]
                        [
                            \textbf{Crisis intervention systems}:\\
                            Suicide prevention dialogue \cite{barak2007emotional};
                            , leaf, draw=c4, line width=0.7mm, edge={c4}
                        ]
                    ]
                    [
                        \textbf{Psychological Educational}\\ \textbf{Interaction}, fill=c4!60, draw=c4, line width=0mm, edge={c4}
                        [
                            \textbf{Mental health education and skill training}:\\
                            Interactive mental health education systems \cite{li2013evaluation,fong2022effects}\\
                            \cite{cai2023cathill,zhang2025ai};\\
                            Social skills training systems \cite{wallace1980review,porayska2018blending};\\
                            Emotion regulation skill training \cite{yu2025multimodal};
                            , leaf, draw=c4, line width=0.7mm, edge={c4}
                        ]
                        [
                            \textbf{Adaptive learning guidance}:\\
                            Intelligent tutoring systems \cite{liu2025lpitutor,wang2025llm,venugopalan2025combining};\\
                            Motivational support systems \cite{meyer2024using,gaeta2025enhancing};\\
                            Metacognitive scaffolding \cite{tomisu2025cognitive,gou2025research,liu2025adaptive};
                            , leaf, draw=c4, line width=0.7mm, edge={c4}
                        ]
                    ]
                ]
            ]
        \end{forest}
    }
    \caption{Comprehensive taxonomy of AI-driven psychological computing tasks.}
    \label{fig: taxonomy_psych}
\end{figure*}

\definecolor{c1}{HTML}{800000} % Deep Red - Classification
\definecolor{c2}{HTML}{C76B1F} % Deep Orange - Regression
\definecolor{c3}{HTML}{1F4E79} % Deep Blue - Structured Relational
\definecolor{c4}{HTML}{B8860B} % Dark Goldenrod - Reasoning and Generation
\definecolor{c5}{HTML}{2E8B57} % Dark Green - Interactive

Building on the formal framework in Section~\ref{sec:formal_formulation}, this taxonomy organizes AI-driven psychology tasks by their computational processing patterns.
It provides mathematical language for describing any psychological computing task through $\mathcal{T} = \langle \mathcal{X}, \mathcal{Y}, \mathcal{F} \rangle$.
Our taxonomy examines how different instantiations give rise to distinct task categories with specific algorithmic requirements.
The structure of output space $\mathcal{Y}$ and computational objective of function $\mathcal{F}$ prove particularly important.

We deliberately avoid organizing by application domains or psychological subdisciplines.
Tasks from different psychological domains often share identical computational requirements, while tasks within the same domain may demand fundamentally different algorithmic approaches.
A depression detection system analyzing social media and a cognitive assessment system processing clinical interviews may employ identical transformer architectures, fine-tuning strategies, and evaluation metrics.
Conversely, within mental health applications, depression severity prediction requires regression while suicide risk detection demands classification, despite addressing related phenomena.

We categorize tasks based on four computational characteristics.
Output space structure indicates whether the task produces discrete categories, continuous values, structured representations, or dynamic content.
Computational objective reveals whether the system performs pattern recognition, numerical prediction, relationship modeling, or content generation.
Model architecture requirements differ across discriminative models, graph neural networks, and generative systems.
Training paradigms vary from supervised learning to structure learning and generative training with reinforcement learning.

This computational lens enables systematic identification of transferable methodological patterns.
A multi-task learning framework for depression severity prediction adapts to anxiety level quantification because both involve regression with similar structures.
A graph attention mechanism for emotion network analysis repurposes for social support network modeling because both involve structured relational learning.
A prompt engineering strategy effective in therapeutic response generation transfers to personalized learning content creation because both require generative capabilities adapting to user context.

We identify four primary task types based on these patterns.
\textbf{Classification tasks} map multimodal psychological data to discrete categorical labels through discriminative pattern recognition.
These tasks operationalize established diagnostic frameworks such as DSM-5, personality taxonomies including the Big Five, and behavioral typologies.
\textbf{Regression tasks} predict continuous numerical values quantifying psychological constructs along graded scales, reflecting that most mental phenomena exist along spectra.
Examples include symptom severity scores, cognitive ability measures, and treatment outcome predictions.
\textbf{Structured relational tasks} model complex interdependencies and temporal dynamics through graph structures, hierarchies, or temporal sequences.
These tasks capture relationships between psychological entities, temporal evolution of mental states, and sequential patterns in clinical processes.
\textbf{Generative and interactive tasks} create personalized psychological content and maintain sustained engagement through dynamic response generation.
These tasks range from single-turn content synthesis to multi-turn conversational interactions.

Each category exhibits distinct computational requirements, model architectures, and evaluation methodologies that transcend specific psychological domains.
Figure~\ref{fig: taxonomy_psych} provides a comprehensive overview of all task types and their representative applications within this taxonomy.
The following sections examine each category in detail, explicating the computational-psychological mapping that makes these tasks tractable for AI systems.

\subsection{Classification Tasks}
\label{sec:3.2_classification}

Classification tasks map multimodal psychological data to discrete categorical labels, operationalizing established diagnostic frameworks such as DSM-5, personality taxonomies including the Big Five, and behavioral typologies.
These tasks encompass mental state recognition, which identifies transient psychological conditions such as depression and anxiety, and behavioral pattern classification, which categorizes stable characteristics such as learning styles and addiction tendencies.

\subsubsection{Mental State Recognition}
\label{sec:mental_state_recognition}

Mental state recognition identifies discrete psychological conditions from observable behavioral manifestations, operationalizing diagnostic frameworks where conditions such as depression, anxiety, and cognitive impairments are categorized based on measurable symptom patterns.

\textcolor{c1}{Affective disorder detection} identifies emotional and mood-related conditions.
Depression detection operates through binary and severity-based classification distinguishing individuals with major depressive disorder from healthy controls while categorizing severity levels \cite{chikersal2021detecting,hussain2025multi}.
Anxiety recognition categorizes individuals into ordinal severity levels based on symptom manifestations \cite{lovibond1995structure,yang2025concise}.
Bipolar disorder classification discriminates mania, hypomanic, and remission states from longitudinal behavioral signals \cite{ringeval2018avec}.

\textcolor{c1}{Cognitive function assessment} categorizes intellectual and processing capabilities through systematic evaluation.
Cognitive decline detection distinguishes individuals with progressive impairment from those with normal aging, identifying early indicators preceding formal diagnosis of mild cognitive impairment or dementia \cite{wang2021development,du2024enhancing}.
Attention deficit detection maps multimodal behavioral features to ADHD diagnostic labels, achieving early identification before formal clinical diagnosis \cite{kim2023machine,garcia2023predicting}.
Autobiographical memory assessment categorizes memory specificity levels, distinguishing episodic details from overgeneral recollections \cite{takano2017unraveling}.

\textcolor{c1}{Personality-based classification} transforms individual characteristics into categorical frameworks through established taxonomies.
Big Five trait recognition categorizes individuals along five personality dimensions through linguistic and behavioral analysis \cite{han2020knowledge,ren2021sentiment,maharjan2025psychometric}.
Attachment style assessment categorizes individuals into attachment security patterns based on interpersonal behavior patterns \cite{gomez2023online}.
Coping style detection distinguishes adaptive and maladaptive stress management strategies based on behavioral responses to stressors \cite{golbeck2016detecting}.

\textcolor{c1}{Mental health screening applications} provide systematic risk stratification through population-level assessment.
Mental health risk assessment identifies individuals requiring intervention through classification of clinical and behavioral data \cite{garriga2022machine,gao2023machine}.
Suicide risk prediction categorizes individuals into risk levels through analysis of linguistic and behavioral patterns \cite{rabani2023detecting}.
Post-traumatic stress disorder recognition operates through classification of symptom patterns and physiological responses \cite{galatzer2017utilization}.

\subsubsection{Behavioral Pattern Classification}
\label{sec:behavioral_pattern_classification}

Behavioral pattern classification transforms complex human behaviors into discrete categorical decisions, encompassing learning styles, social behaviors, and addictive patterns that are evaluated against established categorical frameworks in psychological practice.

\textcolor{c1}{Learning behavior analysis} focuses on educational engagement patterns through multiple dimensions.
Learning style recognition classifies individuals into established pedagogical frameworks through analysis of learning behaviors and interaction patterns \cite{lo2005identification,ozpolat2009automatic,bernard2022improving}.
Learning difficulty diagnosis identifies students with learning disabilities through computational classification of academic performance patterns \cite{wu2008evaluation,nanni2009ensemble}.
Learning motivation classification predicts students' motivational states based on learning behaviors and engagement patterns \cite{babic2017machine}.

\textcolor{c1}{Social behavior assessment} transforms interpersonal dynamics into categorical frameworks through systematic social functioning evaluation.
Social anxiety detection categorizes social interaction patterns into distinct symptom severity levels \cite{salekin2018weakly,ohse2024gpt}.
Interpersonal relation prediction classifies interpersonal dynamics into relationship types along affective and behavioral dimensions \cite{zhang2018facial,gao2021multi}.

\textcolor{c1}{Addiction behavior recognition} identifies and categorizes addictive behaviors through systematic pattern analysis.
Substance dependence detection focuses on binary and multi-class identification of dependencies across psychoactive substances including alcohol, opioids, stimulants, and cannabis through consumption patterns, withdrawal symptoms, and functional impairment indicators \cite{hassanpour2019identifying,rane2021aim,sarker2019machine,afshar2022development,kanan2023intelligent}.
Behavioral addiction classification extends this framework to non-substance addictive behaviors including internet gaming disorder, gambling disorder, and compulsive shopping \cite{king2014cognitive,potenza2019gambling,muller2021object}.

\subsection{Regression Tasks}
\label{sec:3.3_regression}

Regression tasks predict continuous numerical values for psychological phenomena, operationalizing dimensional theories where constructs such as symptom severity, cognitive abilities, and emotional intensities exist along continuous spectra.
These tasks encompass psychological measurement using established clinical instruments, continuous psychological variables quantifying dimensional constructs, and treatment outcome prediction forecasting therapeutic effectiveness.

\subsubsection{Psychological Measurement}
\label{sec:psychological_measurement}

Psychological measurement predicts scores on established clinical assessment instruments with validated continuous scales, enabling quantification of constructs such as depression severity and cognitive abilities that exist along dimensional spectra.

\textcolor{c2}{Severity assessment} quantifies symptom intensity along continuous scales through regression models mapping behavioral indicators to numerical severity scores.
Depression severity assessment predicts continuous symptom scores across established rating scales, enabling fine-grained quantification of depressive symptomatology \cite{haque2018measuring,uddin2022deep}.
Anxiety severity quantification estimates continuous anxiety levels, supporting dimensional assessment of anxiety symptoms \cite{jacobson2020digital,jacobson2022anxiety}.
Stress level assessment predicts continuous stress levels from physiological and behavioral indicators, enabling objective measurement of stress responses across multiple contexts \cite{giannakakis2019review}.

\textcolor{c2}{Cognitive performance prediction} quantifies individual differences in fundamental cognitive abilities through continuous numerical scores.
Intelligence quotient estimation predicts continuous IQ scores along standardized scales, enabling precise quantification of individual differences in general cognitive ability \cite{di2012intelligent}.
Working memory capacity prediction quantifies continuous individual differences in temporary information storage and manipulation abilities, enabling precise measurement of capacity limits across diverse cognitive tasks \cite{cowan2012models}.
Attention control level estimation quantifies continuous individual differences in the ability to voluntarily regulate attention, enabling precise measurement of attentional control capacity across diverse cognitive contexts \cite{unsworth2024individual}.

\subsubsection{Continuous Psychological Variables}
\label{sec:continuous_psychological_variables}

Continuous psychological variables quantify psychological phenomena along dimensional scales, encompassing affective experiences, personality traits, behavioral frequencies, and neuropsychological indicators that manifest as continuous spectra with measurable variations in intensity and magnitude.

\textcolor{c2}{Emotional state quantification} represents systematic measurement of affective experiences along continuous numerical dimensions.
Dimensional emotion assessment quantifies emotional experiences using the circumplex model, which characterizes affect along two continuous dimensions: valence measuring the pleasantness or unpleasantness of emotional states and arousal measuring the activation intensity ranging from low-energy states like calm to high-energy states like excited \cite{mollahosseini2017affectnet,joudeh2024predicting,nomiya2025artificial}.
This dimensional approach enables precise quantification of individual differences in emotional experiences beyond discrete categorical classifications.

\textcolor{c2}{Personality dimension scoring} quantifies individual differences across continuous personality trait dimensions.
Continuous personality trait measurement assesses individual differences across stable behavioral, cognitive, and emotional patterns, enabling precise evaluation along dimensional continua rather than categorical classifications \cite{mccrae1992introduction,bleidorn2019using}.
Temperament dimensions, representing biologically-based individual differences in reactivity and self-regulation, are similarly assessed through continuous scales \cite{rothbart1981development}.

\textcolor{c2}{Behavioral frequency prediction} quantifies the occurrence rates and intensity levels of human behaviors through continuous numerical outputs.
Activity level prediction measures continuous physical activity intensity, and research demonstrates that such prediction can be based on psychological states such as self-efficacy \cite{zhaoyang2017morning}.
Social interaction frequency quantifies interpersonal communication rates as a continuous behavioral indicator predicted by psychological states such as daily perceived stress \cite{dasilva2021daily}.

\textcolor{c2}{Neuropsychological indicators} represent continuous psychological variables quantifying fundamental cognitive processes through precise temporal and performance measurements.
Reaction time prediction measures continuous response latencies, reflecting individual differences in information processing speed and cognitive efficiency \cite{der2006age,bielak2010intraindividual}.
Cognitive processing speed measures continuous efficiency of mental operations across multiple cognitive domains, representing a fundamental neuropsychological indicator of information processing capacity \cite{kail1994processing,hale1994global}.

\subsection{Structured Relational Tasks}
\label{sec:3.4_structured_relational}

Structured relational tasks model complex psychological phenomena through structured outputs including networks, graphs, sequences, and temporal patterns.
Unlike classification tasks that produce discrete category labels or regression tasks that generate continuous scalar values, structured relational tasks capture relationships between entities, temporal dynamics across observations, and hierarchical organizations within psychological data.
These tasks encompass structured relational network modeling that constructs graph-based representations of psychological phenomena and structured temporal and trajectory modeling that captures time-dependent processes through sequential structures.

\subsubsection{Structured Relational Network Modeling}
\label{sec:structured_relational_network}

Structured relational network modeling constructs graph-based representations of psychological phenomena by identifying entities and their relationships.
These tasks output network structures that capture semantic associations, symptom co-occurrences, interpersonal dynamics, and knowledge organization.
Nodes denote psychological entities while edges represent their connections.

\textcolor{c3}{Psychological network construction} builds graph structures representing relationships between psychological entities.
Nodes denote concepts, symptoms, or traits while edges capture their associations, co-occurrences, or dependencies.
Semantic association networks map conceptual relationships in cognitive systems through graphs where nodes represent psychological concepts and edges denote semantic connections derived from word associations or co-occurrence patterns \cite{steyvers2005large,de2019small}.
Symptom networks model psychopathology structure where symptoms serve as nodes and their co-occurrence, temporal precedence, or causal relationships form edges, revealing symptom interactions underlying mental disorders \cite{schmittmann2013deconstructing,bringmann2016assessing}.
Personality trait networks capture individual differences through interconnected trait nodes with edges representing statistical dependencies, revealing the dynamic structure of personality organization \cite{costantini2015state}.

\textcolor{c3}{Social and interpersonal network modeling} constructs graph representations of social relationships and interpersonal interactions in mental health contexts.
Nodes represent individuals while edges capture support relationships, communication patterns, or therapeutic interactions.
Social support networks identify supportive relationships among individuals by extracting mentions of social connections from clinical narratives or online health communities, producing graphs that reveal social isolation or support availability \cite{lu2021development}.
Interpersonal interaction networks model dynamic exchanges between individuals in therapeutic settings through turn-taking patterns, emotional reciprocity, and conversational structures, outputting temporal graphs of interaction dynamics \cite{hale2023hidden}.

\textcolor{c3}{Mental health knowledge graph construction} organizes domain knowledge into structured representations through extraction of entities such as disorders, symptoms, treatments, and risk factors along with their semantic relationships.
These graphs integrate clinical concepts and their interconnections.
Clinical knowledge graphs extract structured information from mental health literature and clinical records through entity and relation identification, constructing comprehensive graphs linking disorders, symptoms, treatments, and outcomes \cite{park2024leveraging,gao2025large}.
Domain-specific knowledge graphs integrate specialized mental health knowledge through connections between psychological concepts and related domains such as nutrition, or social determinants, enabling cross-domain reasoning \cite{dang2023gena,shang2025integrating}.

\subsubsection{Structured Temporal and Trajectory Modeling}
\label{sec:structured_temporal_trajectory}

Structured temporal and trajectory modeling captures time-dependent psychological processes through sequential outputs.
These tasks extract temporal patterns, state transitions, and developmental trajectories from longitudinal psychological data.
The outputs include structured representations of clinical timelines, symptom dynamics, and treatment pathways.

\textcolor{c3}{Temporal event extraction and sequence modeling} identifies time-stamped psychological events and constructs temporal sequences of symptom onset, state transitions, and developmental trajectories.
Symptom onset timeline extraction structures temporal information from clinical notes by identifying when psychological symptoms first emerged, supporting early intervention research in mental health conditions such as psychosis \cite{viani2018time,viani2020temporal,viani2021natural}.
Intraindividual state sequence modeling captures within-person symptom dynamics through inference of hidden psychological states and transition probabilities from longitudinal assessment data, outputting state sequences and transition matrices of individual symptom trajectories \cite{liu2024intraindividual}.

\textcolor{c3}{Clinical pathway and trajectory mining} discovers sequential patterns in treatment processes and patient journeys through extraction of structured pathways from longitudinal clinical data.
Outputs include directed graphs or sequence patterns of typical treatment trajectories.
Treatment pathway discovery identifies common sequences of clinical interventions, service utilization, and patient outcomes from healthcare databases, producing structured representations of treatment processes that reveal standard care patterns and outcome-associated pathways \cite{muyama2024machine}.

\subsection{Generative and Interactive Tasks}
\label{sec:3.5_generative_interactive}

Generative and interactive tasks create personalized psychological content through computational synthesis and dynamic engagement.
These tasks span from single-turn content generation to sustained multi-turn conversational interactions, encompassing personalized content generation that produces tailored therapeutic and educational materials, psychological service interaction that provides mental health support through sustained dialogue, and psychological educational interaction that facilitates adaptive learning through dynamic instruction.

\subsubsection{Personalized Content Generation}
\label{sec:personalized_content_generation}

Personalized content generation creates contextually appropriate psychological materials by synthesizing psychological knowledge with individual characteristics to produce tailored interventions and educational content.
These tasks bridge abstract psychological principles with concrete individual needs while maintaining theoretical validity and practical applicability.

\textcolor{c4}{Customized therapeutic recommendations} generate individualized treatment plans by reasoning about patient characteristics, symptom patterns, and treatment preferences \cite{wood2008adapting,norcross2011psychotherapy,lutz2022prospective,nye2023efficacy}.

\textcolor{c4}{Personalized learning programs} generate individualized educational content by integrating learner characteristics, knowledge levels, and learning preferences to create adaptive instructional materials \cite{wei2025advancement,tu2025empowering}

\subsubsection{Psychological Service Interaction}
\label{sec:psychological_service_interaction}

Psychological service interaction encompasses therapeutic-oriented tasks providing direct mental health support through sustained AI-mediated communication.
These systems maintain continuous engagement while delivering evidence-based psychological interventions adapted to individual user needs.

\textcolor{c4}{Psychological counseling and therapy dialogue} generates therapeutic responses in sustained multi-turn conversations by integrating user emotional states, therapeutic goals, and session context.
Intelligent psychological counseling robots conduct extended therapeutic conversations through adaptive dialogue systems that maintain empathic engagement \cite{spitale2025vita,xie2025psydt}.
Emotional support dialogue recognizes emotional states, validates feelings, and offers comfort through supportive conversational interactions \cite{liu-etal-2021-towards}.
Therapy assistance dialogue guides users through evidence-based therapeutic protocols with structured interventions \cite{na-2024-cbt}.

\textcolor{c4}{Crisis intervention systems} generate immediate adaptive responses to individuals experiencing acute psychological distress through real-time dialogue.
Suicide prevention dialogue engages individuals contemplating self-harm through sustained interactions that provide empathic support while continuously assessing risk and collaboratively developing safety strategies \cite{barak2007emotional}.

\subsubsection{Psychological Educational Interaction}
\label{sec:psychological_educational_interaction}

Psychological educational interaction generates personalized learning experiences through dynamic, bidirectional engagement that continuously adapts to user responses, knowledge levels, and learning progress.
These systems integrate pedagogical principles with real-time assessment to deliver adaptive instruction, practice opportunities, and feedback.

\textcolor{c4}{Mental health education and skill training} generates interactive educational content and practice scenarios by integrating pedagogical principles, user characteristics, and learning objectives.
Interactive mental health education systems generate adaptive content that responds to user engagement.
These systems create web-based games \cite{li2013evaluation}, storytelling narratives \cite{fong2022effects}, emotion-responsive dialogues \cite{cai2023cathill}, and personalized materials \cite{zhang2025ai} tailored to individual learning needs.
Social skills training systems create realistic social scenarios with real-time feedback to facilitate interpersonal competency acquisition through iterative practice \cite{wallace1980review,porayska2018blending}.
Emotion regulation skill training produces guided practice exercises with adaptive feedback to develop emotional competencies through digital interventions \cite{yu2025multimodal}.

\textcolor{c4}{Adaptive learning guidance} provides personalized educational support through continuous interaction that adapts to individual learning trajectories.
Intelligent tutoring systems generate instructional dialogues that adapt teaching strategies to learner understanding, misconceptions, and knowledge gaps \cite{liu2025lpitutor,wang2025llm,venugopalan2025combining}.
Motivational support systems deliver adaptive encouragement through personalized motivational messages that respond to learner achievement patterns and affective states \cite{meyer2024using,gaeta2025enhancing}.
Metacognitive scaffolding guides self-regulated learning through AI-powered interactive prompts that help learners plan, monitor, and reflect on their learning processes \cite{tomisu2025cognitive,gou2025research,liu2025adaptive}.

\section{Datasets and Evaluation Metrics}
\label{sec:datasets}

\subsection{Overview}

High-quality datasets and appropriate evaluation metrics are essential for advancing AI-driven psychological computing.
While Section \ref{sec:background} examined the theoretical challenges inherent to psychological data, this section provides practical guidance for researchers.
We survey representative datasets across task categories, analyze their characteristics, and present evaluation metrics with clear application contexts.
Our goal is to enable informed resource selection and effective evaluation protocol design.

The landscape of psychological computing datasets has evolved through three distinct phases over the past two decades.
Early datasets featured small to medium-scale collections with several hundred participants in controlled settings.
These provided standardized benchmarks with validated measures but limited ecological validity \cite{lucey2010extended}.
The proliferation of social media platforms enabled unprecedented scale, with pioneering work demonstrating feasibility of mental health measurement from naturalistic data \cite{de2013social}.
Standardized shared tasks followed, facilitating method comparison \cite{valstar2013avec,coppersmith2015clpsych}.
These naturalistic datasets offered greater ecological validity but introduced challenges in annotation quality and label reliability.
This prompted methodological innovations such as digital cohort approaches \cite{amir2019mental}.
Most recently, LLMs have transformed the field by enabling sophisticated extraction of psychological signals from unstructured text, facilitating synthetic data generation, and shifting focus from traditional feature engineering to prompt-based approaches \cite{demszky2023using,pandey2024harnessing}.
This evolution reflects an ongoing tension between controlled experimental rigor and naturalistic ecological validity \cite{trull2012clinical}.

Contemporary datasets span diverse modalities including text, audio-visual signals, physiological measurements, and behavioral logs.
Each modality offers complementary insights into human psychology \cite{schmidt2018introducing,khoo2024machine}.
Collection paradigms vary widely.
Clinical settings provide expert annotations.
Social media platforms offer large-scale naturalistic data.
Wearable sensors enable continuous monitoring \cite{sano2018identifying}.
Crowdsourcing platforms support scalable annotation.
The integration of unobtrusive multimodal sensing has enabled researchers to capture psychological states in naturalistic environments, though this introduces challenges in data quality and interpretability \cite{zhou2015tackling}.
Dataset accessibility ranges from publicly available resources to restricted-access clinical data and proprietary institutional datasets.
This creates significant challenges for reproducibility and method comparison.

Psychological datasets exhibit several cross-cutting characteristics that shape their utility for computational research.
Most datasets are cross-sectional rather than longitudinal, limiting the ability to model change over time \cite{ariens2020time,xu2022globem,hawes2023predicting}.
Label quality varies widely, from expert clinical diagnoses to self-disclosed conditions on social media.
This requires careful validation and uncertainty modeling \cite{arseniev2018type,ernala2019methodological,vieira2022self,rutter2023haven}.
Demographic representation remains a critical challenge.
Most datasets over-represent Western, English-speaking, high-income populations \cite{henrich2010weirdest,nielsen2017persistent,rad2018toward,krys2024weird}.
Privacy constraints fundamentally shape the dataset landscape.
Clinical data cannot be publicly released due to regulations such as HIPAA and GDPR \cite{allen2018hipaa,lustgarten2020digital,de2023guide,jurczuk2024consent}.
Unlike computer vision and natural language processing, psychological computing lacks widely adopted standardized benchmarks.
This leads to inconsistent evaluation protocols and limited reproducibility.

\subsubsection{Dataset Organization Framework}

We organize datasets across all task categories according to a unified framework that reflects fundamental tradeoffs in psychological data collection.
This framework progresses from controlled clinical environments to naturalistic ecological contexts, balancing annotation quality against ecological validity and scale.

\paragraph{Clinical and Laboratory Settings.}
Clinical and laboratory datasets provide expert-validated labels through structured assessments and controlled protocols.
Sample sizes typically range from dozens to several hundred participants due to intensive data collection and privacy regulations.
High-quality annotations validated against clinical standards enable precise measurement, though limited sample sizes constrain generalization and structured conditions may not capture real-world spontaneity.

\paragraph{Longitudinal Multi-Site Studies.}
Longitudinal multi-site initiatives combine repeated measurements over extended periods with multi-site aggregation.
Sample sizes range from hundreds to tens of thousands tracked over months to years.
This design enables within-subject trajectory modeling and improves statistical power through demographic diversity, though high costs and participant attrition (typical retention 70-80\% over five years) pose substantial challenges.

\paragraph{Ecological and Naturalistic Settings.}
Ecological datasets capture psychological phenomena in real-world contexts through social media, smartphone sensors, wearable devices, and ecological momentary assessment.
Sample sizes reach thousands to millions of participants, providing high ecological validity through spontaneous behavior.
However, annotation quality varies substantially, with social media self-disclosures showing only 30-50\% concordance with clinical criteria \cite{ernala2019methodological} and passive sensing facing endemic missing data (typical completion 60-70\%).

\paragraph{Specialized and Task-Specific Datasets.}
Specialized datasets address specific research questions through unique combinations of data modalities, annotation schemes, or population characteristics.
These include electronic health records for retrospective analysis, synthetic task environments for cognitive assessment, and domain-specific corpora for narrative analysis.
Sample sizes and data quality vary widely by application domain.

The following sections survey representative datasets for each task category organized according to this unified framework.
This consistent organization facilitates cross-task comparison and enables researchers to identify appropriate resources for their specific objectives.
Table~\ref{tab:datasets} summarizes representative datasets across all four task categories, while Table~\ref{tab:metrics} provides a corresponding overview of evaluation metrics; detailed discussion of each follows in the subsections below.

\begin{table*}[p]
\fontsize{9}{10}\selectfont
\setlength{\tabcolsep}{1.3mm}

\begin{center}
\begin{NiceTabular*}{\textwidth}{@{\extracolsep{\fill}}lccccc}[
    code-before = \rowcolor{blue!15}{1} \rowcolor{gray!15}{3,5,7,9,11,13,15}
]
\Xhline{0.08em}
\multicolumn{6}{c}{\textit{\textbf{Classification}}} \\

\bf Name & \bf Scale & \bf Time & \bf Tasks & \bf Type & \bf Access \\
\hline
AIBL & 1,112 participants & 2009 & AD Progression Classification & Image& \textcolor{orange}{\textbf{!}} \\
ADHD-200 & 973 individuals & 2011 & ADHD subtype  classification & Image & \textcolor{green}{$\checkmark$} \\
DAIC & 621 interviews & 2014 & Distress classification &  Video, Text, Signal & \textcolor{green}{$\checkmark$} \\
CLPsych 2015 & 1,746 participants & 2015 & Depression-PTSD-Control Classification & Text & \textcolor{green}{$\checkmark$} \\
CLPsych 2016 & 1,227 posts & 2016 & Post Severity Triage & Text & \textcolor{green}{$\checkmark$} \\
AMIGOS & 40 participants & 2017 & Affect, personality, mood classification & \makecell[c]{Video, Signal} & \textcolor{green}{$\checkmark$} \\
RSDD & 116,484 participants & 2017 &  Self-harm risk classification & Text & \textcolor{orange}{\textbf{!}} \\
MBTI & 8,675 participants & 2017 & MBTI type classification & Text & \textcolor{green}{$\checkmark$} \\
SMHD & 356,358 participants & 2018 & Mental health prediction & Text & \textcolor{green}{$\checkmark$} \\
\citet{gaur2019knowledge} & 500 participants & 2019 & Suicide risk prediction & Text & \textcolor{green}{$\checkmark$} \\
PANDORA & 17.6M comments & 2021 & Personality  prediction & Text & \textcolor{orange}{\textbf{!}} \\
GLOBEM & 705 participants & 2021 & Mental health prediction & \makecell[c]{Signal} & \textcolor{green}{$\checkmark$} \\
\citet{chikersal2021detecting} & 138 participants & 2021 & Depression status detection & Signal & \textcolor{red}{$\times$} \\
\end{NiceTabular*}

\begin{NiceTabular*}{\textwidth}{@{\extracolsep{\fill}}lccccc}[
    code-before = \rowcolor{blue!15}{1} \rowcolor{gray!15}{3,5,7,9,11}
]
\specialrule{.1em}{.05em}{0.05em}
\multicolumn{6}{c}{\textit{\textbf{Regression}}} \\

\bf Name & \bf Scale & \bf Time & \bf Tasks & \bf Type & \bf Access \\
\hline
IEMOCAP & 7,433 utterances & 2007 & Emotion recognition & Text, Audio, Video& \textcolor{orange}{\textbf{!}} \\
SEMAINE & 95 conversations & 2011 & Emotion recognition & \makecell[c]{Text, Audio, Video} & \textcolor{orange}{\textbf{!}} \\
RECOLA & 46 participants & 2013 & Emotion recognition & \makecell[c]{Audio, Video} & \textcolor{orange}{$\textbf{!}$} \\
StudentLife & 48 participants & 2014 & Depression prediction & Text, Signal & \textcolor{green}{$\checkmark$} \\
\citet{saeb2015mobile} & 28 participants & 2015 & Depression prediction & Signal & \textcolor{red}{$\times$} \\
\citet{canzian2015trajectories} & 28 participants & 2015 & Depression prediction & Signal & \textcolor{red}{$\times$} \\
SEWA & 398 participants & 2019 & Emotion recognition  & Audio, Video & \textcolor{orange}{\textbf{!}} \\
\citet{arevian2020clinical} & 1,101 speech samples & 2020 & Mental health prediction & Audio & \textcolor{red}{$\times$} \\
\citet{stachl2020predicting}	& 624 participants &	2020 & Personality prediction & Signal & \textcolor{green}{$\checkmark$} \\
\citet{shui2023personality} & 80 participants & 2023 & Personality prediction & \makecell[c]{Signal} & \textcolor{green}{$\checkmark$} \\
\end{NiceTabular*}

\begin{NiceTabular*}{\textwidth}{@{\extracolsep{\fill}}lccccc}[
    code-before = \rowcolor{blue!15}{1} \rowcolor{gray!15}{3,5,7}
]
\specialrule{.1em}{.05em}{0.05em}
\multicolumn{6}{c}{\textit{\textbf{Structure}}} \\

\bf Name & \bf Scale & \bf Time & \bf Tasks & \bf Type & \bf Access \\
\hline
MySpace & 3,321 participants & 2012 & Social network  simulation & Text & \textcolor{red}{$\times$} \\
\citet{costantini2015state} & 964 participants & 2015 & Personality network analysis & Text & \textcolor{green}{$\checkmark$} \\
\citet{jin2017emotions} & 14.7M posts & 2017 & Network link prediction & Text & \textcolor{orange}{\textbf{!}} \\
\citet{marian2022network} & 1,072 participants & 2022 & APD network analysis & Text & \textcolor{green}{$\checkmark$} \\
\citet{genois2023combining} & 969 participants & 2023 & Contact networks analysis & Text, Signal & \textcolor{orange}{\textbf{!}} \\
\citet{winter2025mental} & 399 participants & 2025 & Mental health trajectories analysis & Text, Signal & \textcolor{green}{$\checkmark$} \\
\end{NiceTabular*}

\begin{NiceTabular*}{\textwidth}{@{\extracolsep{\fill}}lccccc}[
    code-before = \rowcolor{blue!15}{1} \rowcolor{gray!15}{3,5,7,9,11,13}
]
\specialrule{.1em}{.05em}{0.05em}
\multicolumn{6}{c}{\textit{\textbf{Generation}}} \\

\bf Name & \bf Scale & \bf Time & \bf Tasks & \bf Type & \bf Access \\
\hline
Psych8k & 8,187 QA pairs & 2018 & Counseling response generation & Text & \textcolor{green}{$\checkmark$} \\
Counselchat & 2,775 samples & 2020 & Suicide prevention generation & Text & \textcolor{green}{$\checkmark$} \\
ESConv & 1,053 dialogues & 2021 & Empathic response generation & Text & \textcolor{green}{$\checkmark$} \\
AUGESC & 65K dialogues & 2023 & Empathic response generation & Text & \textcolor{green}{$\checkmark$} \\
SoulChat & 2.3M samples & 2023 & Counseling response generation & Text & \textcolor{green}{$\checkmark$} \\
PsyQA & 22,346 questions & 2023 & Counseling response generation & Text & \textcolor{orange}{\textbf{!}} \\
KokoroChat & 6,589 dialogues & 2024 &Counseling response generation & Text & \textcolor{green}{$\checkmark$} \\
CACTUS & 31,577 dialogues & 2024 & Counseling response generation & Text & \textcolor{green}{$\checkmark$} \\
CPsyCounD & 3,134 dialogues & 2024 & Counseling dialogue generation & Text & \textcolor{orange}{\textbf{!}} \\
PsyDial & 2,382 dialogues & 2025 & Counseling response generation & Text & \textcolor{orange}{\textbf{!}} \\
MESC & 1,019 dialogues & 2025 & Empathic response generation & Text, Audio, Video & \textcolor{green}{$\checkmark$} \\
\bottomrule
\end{NiceTabular*}
\caption{
Overview of psychological computing datasets categorized by task types. 
Access levels: \textcolor{green}{$\checkmark$} Public, \textcolor{orange}{\textbf{!}} Restricted Access, \textcolor{red}{$\times$} Private.
}
\label{tab:datasets}
\end{center}
\end{table*}

\begin{table*}[ht]
\fontsize{9}{10}\selectfont
\setlength{\tabcolsep}{1.3mm}

\begin{center}
\begin{NiceTabular*}{\textwidth}{@{\extracolsep{\fill}}lccc}[
    code-before = \rowcolor{blue!15}{1} \rowcolor{gray!15}{3,5,7,9,11,13,15}
]
\Xhline{0.08em}
\multicolumn{4}{c}{\textit{\textbf{Classification}}} \\

\bf Name & \bf Measures & \bf Range & \bf Application \\
\hline
Accuracy & Proportion of correct predictions & [0,1]$\uparrow$ & Balanced classes \\
Precision & Proportion of correct positive predictions & [0,1]$\uparrow$ & Costly false positives \\
Recall & Proportion of correctly identified positives & [0,1]$\uparrow$ & Costly false negatives \\
F$_1$ & Harmonic mean of precision and recall & [0,1]$\uparrow$ & Equal precision-recall importance \\
F$_\beta$ & Weighted harmonic mean of precision and recall & [0,1]$\uparrow$ & Unequal precision-recall importance \\
Macro F$_1$ & Unweighted average F$_1$ per class & [0,1]$\uparrow$ & Equal class importance \\
Micro F$_1$ & Global F$_1$ from pooled predictions & [0,1]$\uparrow$ & Equal sample importance \\
AUROC & Probability positive ranks higher than negative & [0,1]$\uparrow$ & Threshold-free; imbalance-robust \\
AUPRC & Precision-recall trade-off across thresholds & [0,1]$\uparrow$ & Severe imbalance; positive focus \\
Specificity & Proportion of correctly identified negatives & [0,1]$\uparrow$ & Critical true negatives \\
Balanced Acc. & Average of recall and specificity & [0,1]$\uparrow$ & Equal class consideration \\
MCC & Correlation between predicted and actual labels & [-1,1]$\uparrow$ & Severe class imbalance \\
Cohen's $\kappa$ & Agreement corrected for chance & [-1,1]$\uparrow$ & Human annotation comparison \\
\end{NiceTabular*}

\begin{NiceTabular*}{\textwidth}{@{\extracolsep{\fill}}lccc}[
    code-before = \rowcolor{blue!15}{1} \rowcolor{gray!15}{3,5,7}
]
\specialrule{.1em}{.05em}{0.05em}
\multicolumn{4}{c}{\textit{\textbf{Regression}}} \\

\bf Name & \bf Measures & \bf Range & \bf Application \\
\hline
MAE & Average absolute deviation & [0, $+\infty)$$\downarrow$ & Similar error consequences \\
MSE/RMSE & Average squared deviation & [0, $+\infty)$$\downarrow$ & Large errors costly \\
NRMSE & Normalized average prediction error & [0,1]$\downarrow$ & Cross-dataset comparison \\
Pearson $r$ & Linear association strength & [-1,1] & Linear fit; scale-invariant \\
Spearman $\rho$ & Monotonic relationship strength & [-1,1] & Nonlinear; outlier-robust \\
CCC & Correlation and agreement evaluation & [-1,1]$\uparrow$ & Trend and absolute accuracy \\
\end{NiceTabular*}

\begin{NiceTabular*}{\textwidth}{@{\extracolsep{\fill}}lccc}[
    code-before = \rowcolor{blue!15}{1} \rowcolor{gray!15}{3,5,7,9,11}
]
\specialrule{.1em}{.05em}{0.05em}
\multicolumn{4}{c}{\textit{\textbf{Structure}}} \\

\bf Name & \bf Measures & \bf Range & \bf Application \\
\hline
Exact Match & Proportion of exact target matches & [0,1]$\uparrow$ & Single correct answer \\
Partial Match & Proportion capturing core semantics & [0,1]$\uparrow$ & Multiple acceptable answers \\
Boundary F$_1$ & Overlap between predicted and true spans & [0,1]$\uparrow$ & Partial credit for near-miss \\
Relation Card. & Proportion of correct relationship sets & [0,1]$\uparrow$ & Many-to-many extraction \\
Edge Acc. & Proportion of correctly identified edges & [0,1]$\uparrow$ & Binary network reconstruction \\
Struct. Hamming & Number of differing edges & [0,|E|]$\downarrow$ & Graph structure dissimilarity \\
Weighted Edge Err. & RMS deviation of edge weights & [0,$\infty$)$\downarrow$ & Continuous edge prediction \\
Temporal Stab. & Jaccard similarity between consecutive networks & [0,1]$\uparrow$ & Short-term network stability \\
Centrality Stab. & Rank correlation of node importance & [-1,1]$\uparrow$ & Influential node stability \\
\end{NiceTabular*}

\begin{NiceTabular*}{\textwidth}{@{\extracolsep{\fill}}lccc}[
    code-before = \rowcolor{blue!15}{1} \rowcolor{gray!15}{3,5,7}
]
\specialrule{.1em}{.05em}{0.05em}
\multicolumn{4}{c}{\textit{\textbf{Generation}}} \\

\bf Name & \bf Measures & \bf Range & \bf Application \\
\hline
BLEU & Precision-weighted n-gram overlap & [0,1]$\uparrow$ & Lexical similarity \\
ROUGE-L & Longest common subsequence recall & [0,1]$\uparrow$ & Factual coverage \\
Distinct-n & Proportion of unique n-grams & (0,1]$\uparrow$ & Lexical diversity \\
Perplexity & Model uncertainty for next token & [1,$+\infty$)$\downarrow$ & Language modeling \\
BERTScore & Token-wise contextual similarity & [0,1]$\uparrow$ & Semantic similarity \\
\bottomrule
\end{NiceTabular*}
\caption{
Evaluation metrics for psychological computing tasks categorized by task types.
Symbols: $\uparrow$ indicates higher is better, $\downarrow$ indicates lower is better.
}
\label{tab:metrics}
\end{center}
\end{table*}

\subsection{Classification Datasets and Metrics}

\subsubsection{Representative Datasets}

\paragraph{Clinical and Laboratory Datasets.}
Clinical and laboratory settings provide the foundation for classification benchmarks with expert-validated diagnostic labels.
The DAIC dataset \cite{gratch-etal-2014-distress} contains clinical interviews designed to support diagnosis of psychological distress conditions including depression, PTSD, and anxiety.
It includes PHQ-9 depression assessments \cite{kroenke2002phq} along with other psychological evaluations, and comprises four types of interviews: face-to-face, teleconference, Wizard-of-Oz, and automated agent interactions.
The ADHD-200 dataset \cite{bellec2017neuro} contains resting-state fMRI and structural MRI data collected from 973 individuals across eight international sites.

Multimodal laboratory datasets integrate complementary signals for emotion and personality classification.
The AMIGOS dataset \cite{miranda2018amigos} combines Electroencephalogram (EEG), Electrocardiogram (ECG), Galvanic Skin Response (GSR), frontal HD video, and full-body RGB and depth videos from 40 participants viewing affective stimuli in both individual and group settings.

\paragraph{Longitudinal Multi-Site Studies.}
Large-scale longitudinal initiatives enable investigation of developmental trajectories and disorder progression through repeated assessments over extended periods.
The Australian Imaging, Biomarkers and Lifestyle study \cite{ellis2009australian} has tracked 1,112 participants every 18 months since 2006-2008.

\paragraph{Ecological and Naturalistic Datasets.}
Social media platforms enable large-scale classification through self-disclosed diagnoses.
The CLPsych 2015 Twitter dataset \cite{coppersmith2015clpsych} includes users with self-disclosed depression or PTSD and demographically-matched controls (1,146 training users, 600 test users), pioneering the use of age-gender matching in social media mental health research.
The Reddit Self-reported Depression Diagnosis dataset \cite{yates2017depression} aggregates history posts from 9,210 users with depression diagnoses alongside 107,274 matched controls.
The Self-reported Mental Health Diagnoses (SMHD) dataset \cite{cohan2018smhd} covers nine mental health conditions across 20,406 diagnosed users, enabling multi-class classification and comorbidity analysis.
For personality assessment, PANDORA \cite{gjurkovic2021pandora} provides over 17.6 million Reddit comments from 10,288 users: 1,608 users are labeled with Big Five personality traits and 9,084 with MBTI types.
The MBTI dataset \cite{mbti_kaggle,ontoum2022personality} contains personality type labels for 8,675 users.
The CLPsych 2016 dataset provides 1,227 annotated posts from 65,024 Australian youth mental health forum entries spanning 2012 to 2015, with four urgency levels: green, amber, red, and crisis \cite{milne2016clpsych}.
\citet{gaur2019knowledge} developed a dataset of 500 annotated Reddit users with suicide risk severity levels based on C-SSRS, labeled by four psychiatrists into five categories for early intervention assessment.

Passive sensing captures behavioral manifestations through continuous monitoring.
The GLOBEM dataset \cite{xu2022globem} provides 705 person-years of smartphone and wearable sensor data across four years, capturing location, phone usage, communication patterns, physical activity, sleep, and repeated depression assessments.
The dataset by \citet{chikersal2021detecting} contains 16 weeks of smartphone and Fitbit sensor data from 138 college students, with depression assessments at the beginning and end of the semester.

\subsubsection{Evaluation Metrics}
\label{sec:4.2.2_class_eval}

Classification performance assessment requires careful metric selection, balancing statistical rigor with clinical utility.
We organize metrics by their assumptions about misclassification costs and class distributions, specifying appropriate application contexts for each.

\paragraph{Basic Performance Metrics.}
\textbf{Accuracy} measures the proportion of correct predictions:
\begin{equation*}
\text{Accuracy} = \frac{TP + TN}{TP + TN + FP + FN},
\end{equation*}
where $TP$, $TN$, $FP$, and $FN$ denote true positives, true negatives, false positives, and false negatives.
Accuracy becomes misleading under class imbalance.
In suicide attempt prediction with 2\% base rate, always predicting ``no attempt'' achieves 98\% accuracy despite zero utility.
\textbf{Application context:} Accuracy is appropriate only for balanced datasets where all classes carry equal importance and occur with similar frequencies.

\textbf{Precision} measures the proportion of positive predictions that are correct:
\begin{equation*}
\text{Precision} = \frac{TP}{TP + FP}.
\end{equation*}
Low precision leads to unnecessary follow-up assessments, patient anxiety, and resource waste.
\textbf{Application context:} Precision is critical when false positives are costly, such as in screening programs with limited resources or when false alarms cause significant distress.

\textbf{Recall} measures the proportion of actual positive cases correctly identified:
\begin{equation*}
\text{Recall} = \frac{TP}{TP + FN}.
\end{equation*}
\textbf{Application context:} Recall is essential when false negatives are particularly costly, such as in crisis intervention or screening for severe conditions.

The \textbf{F$_1$ score} provides a harmonic mean of precision and recall:
\begin{equation*}
F_1 = \frac{2 \cdot \text{Precision} \cdot \text{Recall}}{\text{Precision} + \text{Recall}}.
\end{equation*}
\textbf{Application context:} F$_1$ is appropriate when precision and recall carry equal importance.

The \textbf{F$_\beta$ score} adjusts the relative importance of recall versus precision:
\begin{equation*}
F_\beta = (1 + \beta^2) \cdot \frac{\text{Precision} \cdot \text{Recall}}{(\beta^2 \cdot \text{Precision}) + \text{Recall}}.
\end{equation*}
Setting $\beta > 1$ emphasizes recall for screening applications.
Setting $\beta < 1$ prioritizes precision when resources are limited.
\textbf{Application context:} F$_\beta$ enables explicit specification of the precision-recall tradeoff based on domain-specific costs.

\paragraph{Multi-Class Metrics.}
\textbf{Macro F$_1$} computes the metric independently for each class and averages:
\begin{equation*}
F_{1,\text{macro}} = \frac{1}{K} \sum_{k=1}^{K} F_{1,k},
\end{equation*}
where $K$ denotes the number of classes.
\textbf{Application context:} Macro F$_1$ is valuable when rare classes carry equal importance to common ones, such as in personality assessment, where all Big Five dimensions merit balanced performance.

\textbf{Micro F$_1$} aggregates contributions from all classes:
\begin{equation*}
F_{1,\text{micro}} = \frac{2 \sum_{k=1}^{K} TP_k}{\sum_{k=1}^{K} (2 \cdot TP_k + FP_k + FN_k)}.
\end{equation*}
\textbf{Application context:} Micro F$_1$ is appropriate when all individual samples are equally important regardless of their class distribution.
In depression severity classification with imbalanced data where most patients exhibit mild or moderate symptoms, Micro F$_1$ will be dominated by performance on these majority classes.
This makes Micro F$_1$ suitable when the research goal prioritizes overall prediction accuracy across all patients rather than balanced performance across severity levels.

\textbf{Confusion matrices} tabulate predicted versus true class assignments, revealing specific confusion patterns.
\textbf{Application context:} Essential for understanding which classes are confused, guiding targeted improvements.

\paragraph{Threshold-Independent Metrics.}
The Area Under the Receiver Operating Characteristic Curve (\textbf{AUROC}) measures the probability that a randomly selected positive instance receives a higher score than a negative instance:
\begin{equation*}
\text{AUROC} = \int_{0}^{1} \text{TPR}(t) , d(\text{FPR}(t)),
\end{equation*}
where TPR denotes the proportion of actual positives correctly identified and FPR denotes the proportion of actual negatives incorrectly classified as positive at various thresholds $t$.
AUROC ranges from 0.5 (random) to 1.0 (perfect).
\textbf{Application context:} 
AUROC is valuable when optimal decision thresholds vary across clinical contexts.
In suicide risk screening, emergency departments may prioritize detecting all at-risk patients while outpatient clinics may tolerate missing some cases to reduce false alarms.
AUROC provides threshold-independent assessment across these scenarios.
However, AUROC can be misleadingly optimistic under severe class imbalance.

The Area Under the Precision-Recall Curve (\textbf{AUPRC}) focuses on positive class performance:
\begin{equation*}
\text{AUPRC} = \int_{0}^{1} \text{Precision}(r) , dr.
\end{equation*}
\textbf{Application context:} More informative than AUROC for imbalanced datasets.
In suicide risk detection with 1-2\% prevalence, AUPRC provides realistic assessment of clinical utility.

\paragraph{Class-Balance Aware Metrics.}
\textbf{Specificity} measures the proportion of actual negative cases correctly identified:
\begin{equation*}
\text{Specificity} = \frac{TN}{TN + FP}.
\end{equation*}
\textbf{Application context:} Important when correctly identifying negative cases matters, such as ruling out conditions or reducing false alarms.

\textbf{Balanced accuracy} averages recall and specificity:
\begin{equation*}
\text{Balanced}_\text{Acc} = \frac{1}{2}\left(\text{Recall} + \text{Specificity}\right).
\end{equation*}
where recall measures the proportion of actual positives correctly identified and specificity measures the proportion of actual negatives correctly identified. A value of 0.5 represents random guessing even under severe imbalance. 
\textbf{Application context:} Balanced accuracy is appropriate when both classes merit equal consideration despite different frequencies in the population. In anxiety disorder screening where most individuals are healthy, balanced accuracy ensures that model performance on the minority anxious group receives equal weight as performance on the majority healthy group.

\textbf{Matthews Correlation Coefficient (MCC)} measures classification quality by symmetrically incorporating all confusion matrix elements, treating correct predictions and errors equally: 
\begin{equation*}
\text{MCC} = \frac{TP \times TN - FP \times FN}{\sqrt{P_{\text{pred}} \cdot P_{\text{true}} \cdot N_{\text{pred}} \cdot N_{\text{true}}}},
\end{equation*}
where $P_{\text{pred}} = TP + FP$ denotes predicted positives, $P_{\text{true}} = TP + FN$ denotes actual positives, $N_{\text{pred}} = TN + FN$ denotes predicted negatives, and $N_{\text{true}} = TN + FP$ denotes actual negatives.
MCC ranges from -1 for perfect disagreement through 0 for random prediction to +1 for perfect agreement.
\textbf{Application context:} MCC is particularly valuable for binary classification tasks with severe class imbalance where a single comprehensive metric is desired.
In depression detection from social media posts, MCC accounts for the large number of true negatives that accuracy-based metrics may overlook.

\textbf{Cohen's kappa coefficient} measures inter-rater agreement correcting for chance:
\begin{equation*}
\kappa = \frac{p_o - p_e}{1 - p_e},
\end{equation*}
where $p_o$ denotes observed agreement and $p_e$ represents expected agreement by chance.
\textbf{Application context:} Appropriate when assessing agreement between computational predictions and human annotations.

\subsection{Regression Datasets and Metrics}

\subsubsection{Representative Datasets}

\paragraph{Clinical and Laboratory Datasets.}
Clinical settings provide continuous-valued annotations through standardized psychological instruments.
The StudentLife dataset \cite{wang2014studentlife} collected passive sensing from 48 college students over 10 weeks with repeated PHQ-9 assessments measuring depression severity from 0 to 27.
The mental health clinic dataset \cite{arevian2020clinical}  aggregated 1,101 phone calls from 47 patients with serious mental illness, with provider global assessment ratings on a 1-10 scale, demonstrating clinical state tracking through speech-derived features. 

Laboratory environments enable synchronized multimodal collection with continuous affective annotations.
The IEMOCAP dataset \cite{busso2008iemocap} provides 12 hours of dyadic interactions from 10 actors with continuous valence, activation, and dominance ratings on 1 to 5 scales.
The RECOLA dataset \cite{ringeval2013introducing} includes spontaneous interactions from 46 French participants with continuous arousal and valence annotations at 40ms intervals on a scale from negative one to positive one, along with synchronized audio, video, ECG, and EDA recordings.
The SEWA database \cite{kossaifi2019sewa} contains recordings from 398 subjects across six cultures, with continuous valence, arousal, and liking annotations ranging from 0 to 1 for video watching and conversation segments. 

\paragraph{Longitudinal Multi-Site Studies.}
Regression tasks in computational psychology require datasets with continuous measurements of psychological constructs over time. 
However, longitudinal multi-site studies tracking continuous symptom trajectories and quantitative psychological changes remain underrepresented in the current dataset landscape.
Most existing multi-site initiatives focus on categorical diagnostic outcomes rather than continuous measurements. 

Nevertheless, comprehensive datasets with repeated continuous psychological assessments remain scarce, limiting research on symptom progression dynamics, treatment response trajectories, and temporal patterns of psychological functioning. 

\paragraph{Ecological and Naturalistic Datasets.}
Smartphone based passive sensing enables continuous assessment in naturalistic environments.
The location-depression dataset \cite{saeb2015mobile} tracked GPS location and phone usage patterns from 28 participants over two weeks, with baseline depression assessments using PHQ-9 questionnaires. 
Using the MoodTraces mobile application, researchers gathered GPS location data and daily PHQ-8 assessments from 28 participants across an average of 71 days, revealing significant correlations between mobility metrics and depressive mood states \cite{canzian2015trajectories}. 

Wearable devices enable continuous physiological measurement.
\citet{shui2023personality} collected heart rate data from 80 male students wearing Fizzo bracelets across five daily situations over 10 days, with Big Five assessments completed beforehand and daily emotion ratings during recording.
The dataset \cite{stachl2020predicting} collected 30 days of smartphone logs from 624 volunteers with 1,821 behavioral predictors for predicting continuous Big Five personality scores through regression.

\paragraph{Specialized and Task-Specific Datasets.}
Specialized datasets quantify cognitive abilities through carefully designed tasks.
The SEMAINE database \cite{mckeown2011semaine} recorded 150 participants interacting with virtual agents across 959 conversations with continuous annotations of valence, activation, power, expectation, and intensity.

\subsubsection{Evaluation Metrics}

Regression assessment requires metrics quantifying both prediction accuracy and agreement with ground truth.
We organize metrics by their mathematical properties and suitability for different contexts, specifying appropriate applications for each.

\paragraph{Error-Based Metrics.}
\textbf{Mean Absolute Error (MAE)} computes average absolute deviation:
\begin{equation*}
\text{MAE} = \frac{1}{n}\sum_{i=1}^{n}|y_i - \hat{y}_i|,
\end{equation*}
where $y_i$ denotes true value and $\hat{y}_i$ represents prediction.
MAE provides linear penalty treating all deviations equally.
\textbf{Application context:} Appropriate when all errors carry similar consequences, such as personality assessment where small and large errors are equally undesirable.

\textbf{Mean Squared Error (MSE)} and \textbf{Root Mean Squared Error (RMSE)} penalize larger errors through quadratic loss:
\begin{equation*}
\text{MSE} = \frac{1}{n}\sum_{i=1}^{n}(y_i - \hat{y}_i)^2, \quad \text{RMSE} = \sqrt{\text{MSE}}.
\end{equation*}
RMSE returns error in original scale, facilitating interpretation.
\textbf{Application context:} Appropriate when large errors are disproportionately costly, such as severely underestimating suicide risk scores.

\textbf{Normalized RMSE (NRMSE)} enables cross-scale comparison by standardizing RMSE:
\begin{equation*}
\text{NRMSE} = \frac{\text{RMSE}}{\max(y) - \min(y)} \quad \text{or} \quad \frac{\text{RMSE}}{\sigma_y},
\end{equation*}
where the denominator is either the range of observed values or the standard deviation $\sigma_y$ of the outcome variable.
\textbf{Application context:} Essential when comparing models across different constructs measured on different scales, such as comparing depression severity scores with anxiety symptom counts.

\paragraph{Correlation Metrics.}
\textbf{Pearson correlation (r)} measures linear association:
\begin{equation*}
r = \frac{\sum_{i=1}^{n}(y_i - \bar{y})(\hat{y}_i - \bar{\hat{y}})}{\sqrt{\sum{i=1}^{n}(y_i - \bar{y})^2}\sqrt{\sum_{i=1}^{n}(\hat{y}_i - \bar{\hat{y}})^2}}.
\end{equation*}
Pearson correlation ranges from -1 through 0 to +1.
High correlation indicates correct ranking despite systematic bias.
\textbf{Application context:} Assesses how well predictions match the pattern of true scores. Does not measure absolute accuracy and should be used with error metrics such as MAE or RMSE. 
However, insensitive to systematic bias necessitates complementary error-based metrics.

\textbf{Spearman rank correlation ($\rho$)} provides non-parametric alternative robust to outliers:
\begin{equation*}
\rho = 1 - \frac{6\sum_{i=1}^{n}d_i^2}{n(n^2-1)},
\end{equation*}
where $d_i$ is the difference between the rank of predicted value $\hat{y}_i$ and the rank of actual value $y_i$ after sorting both sequences by magnitude.
\textbf{Application context:} Particularly useful for evaluating predictions on ordinal psychological scales where rank order matters more than exact numerical values.

\textbf{The Concordance Correlation Coefficient (CCC)} extends Pearson correlation by penalizing deviations from identity:
\begin{equation*}
\text{CCC} = \frac{2\rho\sigma_y\sigma_{\hat{y}}}{\sigma_y^2 + \sigma_{\hat{y}}^2 + (\mu_y - \mu_{\hat{y}})^2},
\end{equation*}
where $\rho$ denotes Pearson correlation, $\sigma$ represents standard deviation, and $\mu$ denotes mean.
CCC simultaneously evaluates precision (correlation) and accuracy (closeness to identity).
\textbf{Application context:} Particularly valuable for continuous affect recognition where both temporal dynamics and absolute calibration matter for deployment.

\subsection{Structured Relational Datasets and Metrics}

\subsubsection{Representative Datasets}

Structured relational tasks in psychological computing encompass both explicit information extraction from unstructured text and implicit relationship modeling from structured assessments.
While natural language processing has developed extensive annotated corpora for entity and relation extraction in general domains, psychology-specific information extraction datasets remain scarce due to privacy constraints and the complexity of annotating psychological constructs with unclear textual boundaries.
Most existing datasets support relationship modeling through network construction from questionnaires, behavioral observations, and social interactions rather than direct extraction from clinical narratives.

\paragraph{Clinical and Laboratory Datasets.}
This dataset contains binary responses from 718 community participants and 354 psychiatric inpatients across seven avoidant personality disorder diagnostic criteria. The data structure supports symptom co-occurrence analysis to identify which symptom pairs frequently appear together \cite{marian2022network}.
The dataset includes 964 participants who completed the HEXACO-60 questionnaire, with 60-item responses aggregated into 24 personality facets for network analysis of trait relationships \cite{costantini2015state}.

The dataset tracks 969 participants across four academic conferences using the SocioPatterns platform to record face-to-face proximity events every 20 seconds, paired with personality questionnaires to yield structured tuples of timestamped contact pairs and individual trait profiles \cite{genois2023combining}.

\paragraph{Longitudinal Multi-Site Studies.}
The Corona Health Adolescent Study provides time-series tuples pairing timestamps with mental health assessments from 399 adolescents across 637 follow-up responses during COVID-19, capturing anxiety, depression, and mobility patterns from July 2020 to December 2023 \cite{winter2025mental}.

\paragraph{Ecological and Naturalistic Datasets.}
An emotion dataset from LiveJournal contains 14.7 million mood-labeled posts with network tuples capturing 1.13 million mutual friendships and 14.1 million directed follower ties across 10 years \cite{jin2017emotions}.
The MySpace dataset \cite{Tadic2017} captures two months of interactions from 3,321 users, supporting extraction of directed emotional influence relationships.

\subsubsection{Evaluation Metrics}

Structured relational tasks in psychological computing encompass two evaluation paradigms: tuple-based metrics for explicit relationship extraction from text, and network-based metrics for implicit relationship modeling from structured data.
The former evaluates whether systems correctly extract and categorize psychological relationships mentioned in clinical notes, therapy transcripts, or research literature.
The latter assesses the accuracy of the inferred networks that represent symptom co-occurrence, personality trait associations, or social connections.

\paragraph{Tuple-Based Extraction Metrics.}
We denote a relation tuple as $(e_1, r, e_2)$ where $e_1, e_2$ are entity spans and $r$ is the relation type.
Let $T_{true}$ and $T_{pred}$ represent the sets of ground truth and predicted tuples, respectively.
\textbf{Basic metrics} such as precision, recall, and F$_1$ scores apply when treating complete tuples as classification units.
These metrics follow the definitions in Section~\ref{sec:4.2.2_class_eval} and are not repeated here.
Below, we focus on metrics specific to structured extraction tasks.

\textbf{Exact match accuracy} requires perfect alignment of all tuple components:
\begin{equation*}
\text{Exact Match} = \frac{|T_{pred} \cap T_{true}|}{|T_{true}|},
\end{equation*}
where tuples must match entity spans, entity types, and relation labels simultaneously.
\textbf{Application context:} Suitable for extracting structured relationships from psychological assessments, such as identifying that "rumination'' (cognitive pattern entity) "predicts'' (relation) "depressive symptoms'' (symptom entity) in therapy notes.
Works well when psychological constructs have standardized terminology and clear relational boundaries.
Inappropriate for ambiguous emotional states or culturally variable symptom expressions where boundary disagreement reflects construct complexity rather than extraction error.

\textbf{Partial match accuracy} allows entity boundary tolerance:
\begin{equation*}
\text{Partial Match} = \frac{|T_{pred} \cap_{overlap} T_{true}|}{|T_{true}|},
\end{equation*}
where $\cap_{overlap}$ denotes tuples with overlapping entity spans and matching relation labels.
\textbf{Application context:} Appropriate for psychological constructs with fuzzy boundaries or variable linguistic expressions, such as extracting symptom descriptions from patient narratives or personality trait mentions from open-ended interviews.
Less suitable for standardized assessments requiring exact terminology alignment, such as structured questionnaire coding or predefined behavioral category systems.

\textbf{Boundary-aware F$_1$ score} penalizes span mismatches proportionally:
\begin{equation*}
\text{F$_1$}_{boundary} = \frac{2 \cdot P_{boundary} \cdot R_{boundary}}{P_{boundary} + R_{boundary}},
\end{equation*}
where $P_{boundary}$ and $R_{boundary}$ incorporate Jaccard similarity between predicted and true entity spans.
\textbf{Application context:} Suitable for extracting multi-word psychological terms from client narratives or therapy transcripts where partial matches retain interpretive value.
Not appropriate for binary classification tasks like diagnostic screening or risk assessment requiring exact categorical labels.

\textbf{Relation cardinality accuracy} evaluates whether models extract complete relationship sets:
\begin{equation*}
    \frac{|\{(e_1, R) : T_{pred}(e_1, R) = T_{true}(e_1, R)\}|}{|\{(e_1, R) \in T_{true}\}|},
\end{equation*}
where $T(e_1, R) = \{e_2 : (e_1, R, e_2) \in T\}$ denotes all entities related to $e_1$ through relation $R$.
\textbf{Application context:} Essential when completeness matters, such as extracting all symptoms associated with a mental disorder or all risk factors linked to a psychological condition.
Less relevant for tasks requiring only existence verification of relationships.

\paragraph{Network Prediction Metrics.}
We denote a psychological network as $G = (V, E, W)$ where $V$ is the set of nodes representing psychological constructs, $E$ is the set of edges representing relationships, and $W$ is the weight matrix.
Let $A$ denote the adjacency matrix where $A_{ij} = 1$ if $(i,j) \in E$.

\textbf{Edge accuracy} evaluates relationship identification:
\begin{equation*}
\text{Edge Accuracy} = \frac{|E_{pred} \cap E_{true}|}{|E_{true}|},
\end{equation*}
where $E_{pred}$ and $E_{true}$ are predicted and ground truth edge sets.
\textbf{Application context:} Suitable when gold-standard networks exist, such as comparing automated extraction against expert-annotated symptom co-occurrence or validating against established theoretical models.
Inappropriate when ground truth is ambiguous or when false positives and false negatives have asymmetric costs, requiring precision-recall trade-off analysis instead.

\textbf{Structural Hamming distance} quantifies network dissimilarity:
\begin{equation*}
d_H(G_1, G_2) = |E_1 \triangle E_2| = |E_1 \setminus E_2| + |E_2 \setminus E_1|,
\end{equation*}
where $\triangle$ denotes symmetric difference between edge sets.
\textbf{Application context:} Appropriate for comparing network structures across groups or conditions when edge presence is binary.
Less suitable for weighted networks where edge strength differences matter more than presence-absence, or when networks have different node sets requiring alignment before comparison.

\textbf{Weighted edge error} assesses relationship strength prediction:
\begin{equation*}
\text{RMSE}_{edge} = \sqrt{\frac{1}{|E|}\sum_{(i,j) \in E}(w_{ij}^{pred} - w_{ij}^{true})^2},
\end{equation*}
where $w_{ij}$ represents edge weights from matrix $W$.
\textbf{Application context:} Essential for networks where relationship strength informs intervention prioritization, such as symptom networks with weighted connections indicating co-activation probability.
Less relevant for binary presence-absence networks or when weight estimation uncertainty is high relative to true weight variation.

\paragraph{Temporal Dynamics Metrics.}
We denote a temporal network sequence as $\{G_t\}_{t=1}^T$ where $G_t = (V, E_t, W_t)$ represents the network at time $t$.

\textbf{Temporal network stability} assesses structural consistency:
\begin{equation*}
J(G_t, G_{t+1}) = \frac{|E_t \cap E_{t+1}|}{|E_t \cup E_{t+1}|},
\end{equation*}
where the Jaccard index ranges from 0 to 1.
\textbf{Application context:} Appropriate for distinguishing trait-like stable structures from state-dependent dynamic networks, such as comparing personality networks across contexts versus mood networks across time.
Unsuitable for detecting gradual network evolution, where change point methods are more informative, or for very short time series, where stability estimates lack reliability.

\textbf{Centrality stability} evaluates temporal consistency of node importance using Spearman rank correlation:
\begin{equation*}
\rho_t = \text{corr}(\text{rank}_t(C), \text{rank}_{t+1}(C)),
\end{equation*}
where $C$ denotes centrality measures and $\text{rank}_t$ assigns rank positions at time $t$.
Unlike prediction accuracy metrics that compare predicted versus true values, this metric assesses whether influential nodes maintain their relative importance across consecutive time points.
\textbf{Application context:} Useful for identifying stable core symptoms in depression networks across treatment phases or persistent central traits in personality networks across life transitions.
Less informative when absolute centrality magnitudes are clinically critical or when network composition changes substantially between time points.

\subsection{Generative Interactive Tasks}

\subsubsection{Representative Datasets}

\paragraph{Clinical and Laboratory Datasets.}
Clinical recording datasets capture authentic therapeutic interactions through professional counseling sessions.
The Psych8k dataset \cite{liu2023chatcounselor} transforms 260 real counseling audio recordings into 8,187 query-answer pairs with specialized counseling skill evaluation.
KokoroChat \cite{qi-etal-2025-kokorochat} provides 6,589 conversations with 600,939 utterances collected through role-playing by 480 trained counselors.
PsyDial \cite{qiu-lan-2025-psydial} reconstructs 2,382 conversations totaling approximately 7 million tokens from real client-counselor dialogues \cite{li-etal-2023-understanding} through privacy-preserving four-stage method: Retrieve, Mask, Reconstruct, and Refine. 
Multimodal extensions integrate audio and visual modalities alongside text to capture comprehensive therapeutic communication.
MESC \cite{chu2025towards} comprises 1,019 dialogues and 28,762 utterances from a psychotherapy television series. It includes text, audio, and video data annotated with emotions and therapeutic strategies, supporting four tasks: emotion recognition, strategy prediction, system emotion prediction, and response generation.

\paragraph{Longitudinal Multi-Site Studies.}
Longitudinal therapeutic dialogue datasets tracking individual therapeutic relationships over extended treatment periods remain critically scarce in the current landscape.
While some clinical record datasets capture repeated crisis events, they differ fundamentally from true longitudinal therapeutic dialogue data that would track continuous therapeutic relationships, alliance development, and intervention adaptation over time.
Electronic health record datasets from UK mental health trusts provide longitudinal crisis monitoring but focus on discrete crisis events rather than continuous therapeutic processes.
However, these datasets primarily capture single-session crisis interventions rather than multi-session therapeutic relationships.

The absence of true longitudinal therapeutic dialogue datasets represents a critical gap for understanding therapeutic alliance development, intervention adaptation over time, and long-term treatment effectiveness.
Such datasets would require tracking individual client-therapist dyads across multiple sessions while maintaining privacy protections, posing substantial ethical and logistical challenges.
This limitation fundamentally constrains research on how therapeutic relationships evolve and how AI systems might support sustained therapeutic engagement.

\paragraph{Ecological and Naturalistic Datasets.}
Crowdsourcing platforms enable scalable therapeutic dialogue collection through structured protocols.
The ESConv dataset \cite{liu2021towards} contains 1,053 conversations with 31,410 utterances collected through helper and help-seeker role assignments.
PsyQA \cite{sun-etal-2021-psyqa} aggregates 22,346 questions with 56,063 answers from the Yixinli Chinese counseling platform, requiring restricted access through user agreements.
CounselChat \cite{bertagnolli2020counsel} provides publicly available question-answer pairs where licensed US therapists respond to anonymous mental health concerns.

\paragraph{Specialized and Task-Specific Datasets.}
Large language model synthesis has emerged as a transformative approach for creating therapeutic dialogues at scale while incorporating domain expertise.
The CACTUS dataset \cite{lee2024cactus} generates 31,577 conversations with 995,512 utterances based on cognitive behavioral therapy frameworks.
AUGESC \cite{zheng2023augesc} expands to 65,000 sessions with 1,738,000 utterances through fine-tuned GPT-J completing dialogues from EmpatheticDialogues posts.
SoulChat \cite{chen-etal-2023-soulchat} creates 2,300,248 multi-turn empathy dialogues by converting single-turn psychological counseling Q\&A pairs into multi-turn conversations via ChatGPT with empathy-constrained prompts, followed by manual proofreading.
CPsyCoun \cite{zhang-etal-2024-cpsycoun} constructs three datasets: 
(1) CPsyCounR contains 3,134 high-quality psychological counseling reports covering 9 counseling topics and 7 classic counseling schools;
(2) CPsyCounD reconstructs 3,134 multi-turn dialogues from CPsyCounR through a two-stage Memo2Demo method incorporating psychological supervisor and counselor roles;
(3) CPsyCounE provides 45 carefully selected cases as an evaluation benchmark for multi-turn psychological counseling dialogues.

\subsubsection{Evaluation Metrics}

Generative interactive evaluation requires metrics addressing both linguistic quality and therapeutic appropriateness.
Unlike classification and regression metrics that focus on prediction accuracy, generative interactive metrics must assess response quality, therapeutic validity, and interaction dynamics.
We organize metrics by their computational focus and psychological validity requirements, specifying appropriate applications for each.

\paragraph{Automatic Generation Quality Metrics.}
\textbf{BLEU} measures n-gram overlap between generated and reference responses:
\begin{equation*}
\text{BLEU} = BP \cdot \exp\left(\sum_{n=1}^{N} w_n \log p_n\right),
\end{equation*}
where $p_n$ denotes n-gram precision, $w_n$ represents weights, and $BP$ is the brevity penalty that penalizes overly short outputs.
\textbf{Application context:} Appropriate for measuring surface-level lexical similarity in psychoeducational content generation but may not capture therapeutic effectiveness in counseling dialogues where multiple valid responses exist.

\textbf{ROUGE} is a family of metrics including n-gram overlap (ROUGE-N) and longest common subsequence matching (ROUGE-L).
The ROUGE-L variant computes:
\begin{equation*}
\text{ROUGE-L} = \frac{(1 + \beta^2) R_{\text{lcs}} P_{\text{lcs}}}{\beta^2 R_{\text{lcs}} + P_{\text{lcs}}},
\end{equation*}
where $R_{\text{lcs}}$ and $P_{\text{lcs}}$ denote recall and precision of longest common subsequence.
\textbf{Application context:} Valuable for evaluating content coverage in psychological assessment reports and counseling summaries where comprehensive information inclusion matters.

\textbf{Perplexity} measures model confidence through negative log-likelihood:
\begin{equation*}
\text{PPL} = \exp\left(-\frac{1}{N}\sum_{i=1}^{N} \log P(w_i|w_{<i})\right),
\end{equation*}
where $P(w_i|w_{<i})$ represents conditional probability of word $w_i$ given preceding words $w_{<i}$.
\textbf{Application context:} Useful for comparing model architectures and training procedures but does not directly measure therapeutic quality or appropriateness.

\textbf{Distinct-n} quantifies lexical diversity through unique n-gram ratios:
\begin{equation*}
\text{Distinct-n} = \frac{|\text{unique n-grams}|}{|\text{total n-grams}|}.
\end{equation*}
\textbf{Application context:} Important for therapeutic dialogues where repetitive responses reduce engagement and may signal insufficient personalization to client concerns.

\textbf{BERTScore} measures semantic similarity using BERT embeddings, which are numerical representations of text meaning generated by neural network models.
The metric computes precision $P_{\text{BERT}}$ and recall $R_{\text{BERT}}$ through maximum cosine similarity between generated and reference token embeddings, then combines them into F$_1$ score:
\begin{equation*}
\text{BERTScore}_{\text{F}_1} = \frac{2 \cdot P_{\text{BERT}} \cdot R_{\text{BERT}}}{P_{\text{BERT}} + R_{\text{BERT}}}.
\end{equation*}
\textbf{Application context:} Captures semantic equivalence in therapeutic dialogue where different phrasings convey similar empathy, such as ``I understand your struggle'' versus ``I hear how difficult this is for you''. Useful for evaluating mental health chatbots where response quality depends on emotional tone rather than exact wording.

\paragraph{Counseling Skills Assessment.}
Beyond automatic metrics, human evaluation assesses therapeutic competence through multiple counseling skill dimensions typically rated on five-point Likert scales where 1 indicates very poor performance and 5 represents excellent demonstration.
Professional raters or trained evaluators examine dialogue quality across complementary aspects.

Core therapeutic skills include \textbf{Empathy} for understanding and validating client emotions, \textbf{Active Listening} for attentiveness to verbal and nonverbal cues, and \textbf{Non-judgmental Attitude} for creating safe environments that encourage disclosure.

Exploratory techniques encompass \textbf{Open-ended Questioning} to encourage expansive client responses, \textbf{Issue Clarification} to resolve ambiguities through specific inquiries, and \textbf{Encouraging Self-Exploration} to facilitate client reflection on emotions and behaviors.

Intervention strategies comprise \textbf{Cognitive Restructuring} to challenge distorted thought patterns, \textbf{Guided Questioning} to focus on specific therapeutic goals, and \textbf{Overall Score} to provide holistic assessment of comprehensive counselor competence.

\textbf{Application context:} These dimensions capture therapeutic appropriateness that automatic metrics fail to measure.
Human evaluation remains essential for validating whether generated responses demonstrate clinically sound counseling practices, maintain ethical boundaries, and foster effective therapeutic relationships.
Raters provide both numerical scores and qualitative reasoning to identify specific strengths and deficiencies in generated therapeutic interactions.

\paragraph{Psychological Validity Metrics.}
While counseling skills assess individual response quality, psychological validity metrics evaluate overall therapeutic effectiveness across complete interactions or treatment courses. These measures validate whether systems produce clinically meaningful outcomes rather than merely appropriate single responses \cite{qiu-lan-2025-psydial}. 
\textbf{Cognitive Therapy Rating Scale (CTRS)} evaluates CBT-specific competencies including understanding, interpersonal effectiveness, collaboration, pacing and efficient use of time, guided discovery, focusing on key cognitions, strategy for change, application of cognitive-behavioral techniques, homework, and collaboration.
Professional therapist ratings assess adherence to evidence-based techniques.
\textbf{Application context:} Critical for CBT systems where adherence to structured therapeutic protocols determines treatment effectiveness and distinguishes evidence-based intervention from generic support.

\textbf{Working Alliance Inventory (WAI)} assesses therapeutic relationship quality through three dimensions: goal agreement (shared understanding of treatment objectives), task agreement (consensus on therapeutic activities), and emotional bond (mutual trust and attachment).
\textbf{Application context:} Measures whether AI systems establish collaborative relationships that predict treatment outcomes, as therapeutic alliance consistently predicts effectiveness across treatment modalities.

\textbf{Positive and Negative Affect Schedule (PANAS)} quantifies emotional state changes through self-report scales measuring positive affect (enthusiastic, interested, determined) and negative affect (distressed, upset, scared).
\textbf{Application context:} Evaluates whether therapeutic interactions produce intended emotional changes, with effective interventions typically increasing positive affect and decreasing negative affect over time.

\paragraph{Application-Specific Considerations.}
Metric selection depends critically on the specific generative interactive application.
\textbf{Crisis intervention systems} prioritize safety metrics and rapid response appropriateness over linguistic sophistication, as immediate risk assessment and appropriate escalation matter more than conversational quality.
\textbf{Psychoeducational interactions} emphasize informativeness and comprehension alongside engagement, requiring balance between accessibility and content depth.
\textbf{Long-term therapeutic relationships} require longitudinal evaluation tracking alliance development, sustained engagement over multiple sessions, and progressive movement toward treatment goals.
Comprehensive evaluation protocols combine automatic metrics for scalable assessment with human evaluation for psychological validity.
Linguistic quality alone provides insufficient evidence for therapeutic system deployment.
Systems must demonstrate both technical competence and therapeutic appropriateness through multi-faceted evaluation addressing generation quality, psychological validity, interaction dynamics, and safety considerations.

\section{Methods}
\label{sec:methods}

This section examines computational approaches for AI-driven psychological tasks across two methodological paradigms: PLM-based methods and LLM-based methods.

We organize methods along two dimensions.
The methodological dimension distinguishes PLM-based methods that require task-specific fine-tuning from LLM-based approaches that enable prompt-based adaptation.
The task-oriented dimension follows the taxonomy in Section~\ref{sec:taxonomy}.
Classification and regression tasks share encoding architectures but differ in output layers.
Structured relational tasks model interdependencies through graph-based or temporal architectures.
Generative interactive tasks synthesize contextual responses through sequence-to-sequence frameworks.

Three fundamental technical shifts distinguish these paradigms.
First, the training paradigm evolved from supervised fine-tuning \cite{devlin2019bert,liu2019roberta} requiring thousands of labeled examples to few-shot or zero-shot prompting \cite{brown2020language,kojima2022large} that reduces annotation requirements.
Second, architectural design shifted from specialized components such as graph neural networks for relational tasks \cite{kipf2016semi,velivckovic2017graph} and hierarchical encoders for long documents \cite{yang2016hierarchical} to general-purpose transformers \cite{vaswani2017attention} adapted through natural language instructions \cite{ouyang2022training,wei2021finetuned}.
Third, model scale increased from 110M to 340M parameters in PLMs \cite{devlin2019bert} to billions or trillions in LLMs \cite{brown2020language,achiam2023gpt}.
This scale enables emergent capabilities like chain-of-thought reasoning \cite{wei2022chain,kojima2022large} that fundamentally alter feasible approaches.

\begin{figure*}[t]
    \centering
    \includegraphics[width=\linewidth]{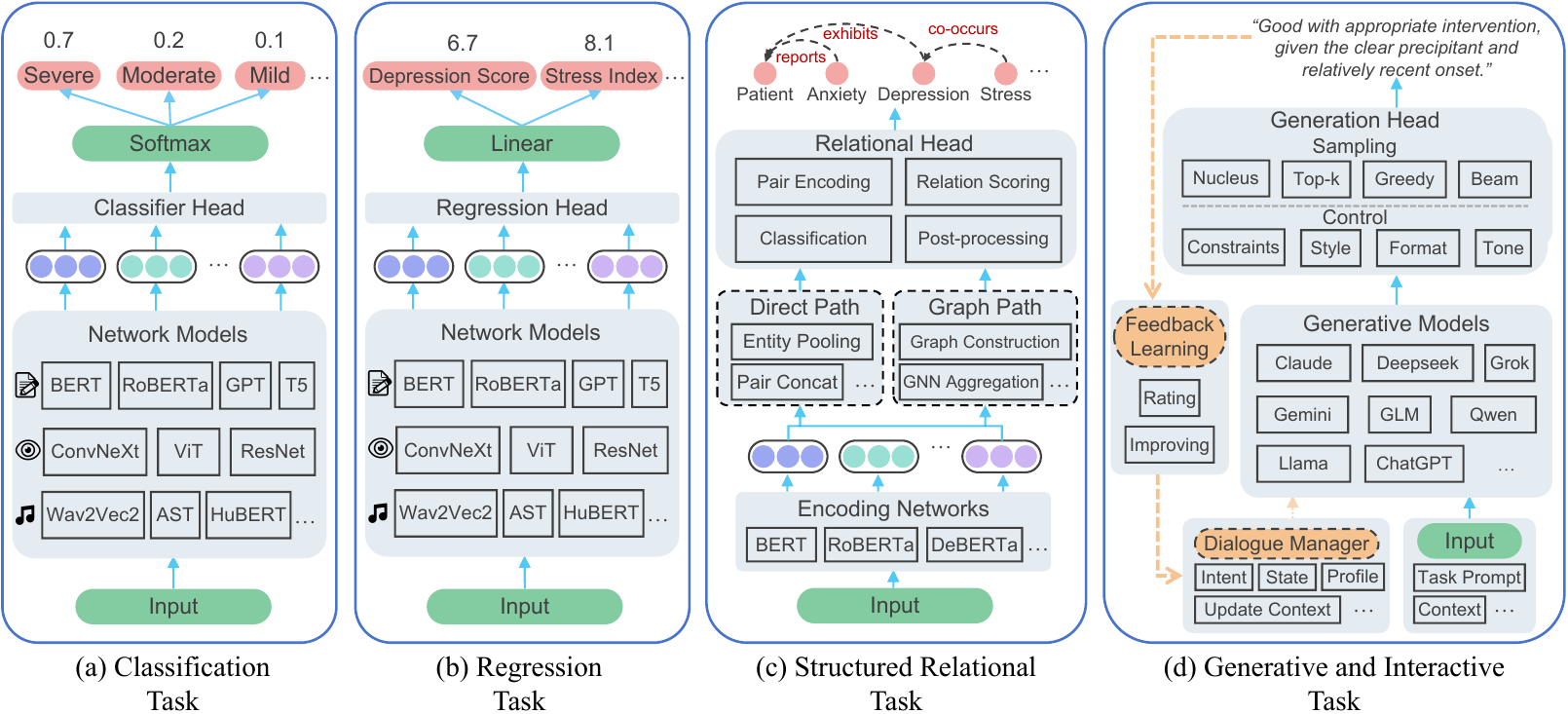}
    \caption{Computational architectures for AI-driven psychological tasks. The figure illustrates technical implementations across four task categories: (a) Classification tasks employ discriminative models with softmax heads for discrete psychological state identification, (b) Regression tasks utilize continuous output layers for quantitative psychological measurement, (c) Structured relational tasks leverage graph-based architectures for modeling construct interdependencies, and (d) Generative interactive tasks integrate content generation subsystems (white components) with dialogue management modules (orange components) for dynamic user engagement. All architectures support multimodal inputs through specialized encoders for text (BERT, RoBERTa, GPT, T5), vision (ViT, ResNet, ConvNeXt), and audio (Wav2Vec2, AST, HuBERT).}
    \label{fig:computational_method}
\end{figure*}

Figure~\ref{fig:computational_method} illustrates how these computational paradigms instantiate across the four task categories.
Classification tasks (Figure~\ref{fig:computational_method}(a)) employ discriminative architectures with softmax output layers, processing multimodal inputs through specialized encoders (BERT/RoBERTa for text, ViT/ResNet for vision, Wav2Vec2/HuBERT for audio) before producing discrete probability distributions over psychological states.
Regression tasks (Figure~\ref{fig:computational_method}(b)) share the same encoding infrastructure but replace classification heads with linear regression layers for continuous psychological measurement.
Structured relational tasks (Figure~\ref{fig:computational_method}(c)) extend beyond instance-level prediction through graph construction and GNN aggregation modules that model interdependencies between psychological constructs.
Generative interactive tasks (Figure~\ref{fig:computational_method}(d)) integrate two complementary subsystems: content generation components (shown in white) that synthesize responses through large generative models with controllable sampling strategies, and dialogue management components (shown in orange) that maintain conversational state through intent recognition, context tracking, and dynamic profile updating for sustained multi-turn engagement.

Section~\ref{sec:plm_based} examines PLM-based solutions including fine-tuning strategies, domain-adaptive pretraining, and architectural adaptations.
Section~\ref{sec:llm_based} explores LLM-based approaches covering prompt engineering, instruction tuning, and in-context learning.
Within each paradigm, we organize methods by task category to highlight both paradigm-specific innovations and transferable patterns across psychological applications.

\subsection{PLM-based Methods}
\label{sec:plm_based}

PLM-based methods typically contain 110M to 340M parameters and require substantial labeled data for task-specific fine-tuning.
The core challenge lies in bridging general language understanding with domain-specific psychological constructs through effective adaptation strategies.

\subsubsection{Classification and Regression Methods}
\label{sec:plm_classification_regression}

Classification and regression tasks in psychological computing share common encoding architectures but differ in their output formulations.
Classification tasks map inputs to discrete psychological categories through softmax layers, while regression tasks predict continuous severity scores or trait measurements through linear output layers.
Both benefit from similar feature extraction strategies and domain adaptation techniques.

\paragraph{Single-Modal Text Methods}

\textbf{Fine-tuning Strategies.}
Standard fine-tuning directly applies pre-trained models to psychological tasks through supervised learning on labeled datasets.
Suicide risk detection employs masked language modeling with suicide-oriented dictionaries to enhance sensitivity to risk-related expressions \cite{cao-etal-2019-latent}.
Personality trait detection extracts contextualized embeddings by concatenating BERT's last four layers and combines them with psycholinguistic features, then feeds the combined representations to a Bagged SVM classifier to capture both semantic content and linguistic style \cite{kazameini2020personality}.
Depression screening uses BERT-based extractive summarization to filter relevant posts from lengthy social media histories \cite{zogan2021depressionnet}, while mental illness classification fine-tunes RoBERTa for multi-class diagnostic categorization \cite{murarka2021classification}.
The 512-token limit of standard transformers poses challenges for long documents common in psychological applications.
Hierarchical encoding addresses this for structured inputs, where Sentence-RoBERTa encodes individual dialogue turns and a second-level encoder processes turn sequences to capture conversational dynamics for predicting symptom severity scores\cite{milintsevich2023towards}.
Personality prediction employs sliding windows that segment texts into overlapping chunks processed by RoBERTa, with an RNN aggregating chunk representations to capture document-level patterns \cite{hussain2025personality}.
Fine-tuned BERT models enable accurate depression assessment in low-resource languages, as demonstrated by MADRS score prediction from German interview transcripts \cite{weber2025using}.

\textbf{Domain-Adaptive Pretraining.}
General pre-trained models lack exposure to psychological terminology and mental health discourse patterns.
Continued pre-training on domain-specific corpora addresses this vocabulary and discourse gap.
ClinicalBERT focuses on electronic health records to capture clinical documentation patterns \cite{huang2019clinicalbert}, while BioBERT specializes in biomedical publications for research-oriented applications \cite{lee2020biobert}.
PsychBERT combines biomedical literature with social media posts to bridge formal clinical language and everyday mental health discussions \cite{vajre2021psychbert}.
MentalBERT and MentalRoBERTa undergo additional pre-training on Reddit mental health communities to learn both clinical terminology and colloquial expressions of psychological distress \cite{ji2022mentalbert}.
CASE-BERT pre-trains on psychology textbooks to acquire structured expert knowledge and theoretical frameworks \cite{harne2024case}.

\textbf{Multi-Task Learning.}
Multi-task learning exploits correlations between psychological constructs by training shared representations across related tasks.
Pre-trained language models enable simultaneous screening of multiple mental illnesses such as major depressive disorder and post-traumatic stress disorder, where shared BERT encoders capture common linguistic patterns while task-specific layers differentiate disorder-specific symptoms \cite{shrestha2024multi}.
Opinion-enhanced BERT architectures jointly model sentiment classification and mental health status detection by integrating opinion embeddings with contextual representations, using hybrid CNN-BiGRU layers to capture both local and sequential features for improved emotional state assessment \cite{hossain2025multi}.

\textbf{Few-Shot Learning.}
Limited labeled data in clinical psychology motivates few-shot learning approaches.
Data augmentation techniques expand training sets through transformations that preserve semantic content.
Easy Data Augmentation applies random insertion, deletion, swap, and synonym replacement operations to generate synthetic training examples \cite{wei2019eda}.
Conditional BERT generates contextually appropriate augmentations by masking tokens and sampling from predicted distributions \cite{wu2019conditional}.
Back-translation creates paraphrases through round-trip translation between languages, introducing lexical variation while maintaining semantic equivalence \cite{sennrich2016improving}.
These augmentation strategies have been successfully applied to mental health classification tasks, where combining multiple techniques addresses data scarcity challenges in clinical datasets \cite{ansari-etal-2021-data}.

\paragraph{Multimodal Fusion Methods}
Multimodal approaches integrate complementary signals from text, audio, video, and physiological sensors to capture the multifaceted nature of psychological states.
Depression detection combines BERT text embeddings processed through a CNN-LSTM architecture with deep spectrum acoustic features extracted via a Gated CNN-LSTM model, where the gating mechanism selectively filters acoustic representations prior to feature-level concatenation \cite{rodrigues2019multimodal}.
Depression severity assessment employs adaptive fusion strategies where modalities pass through both shared encoders that capture common patterns and private encoders that preserve modality-specific information, with adaptive fusion modules dynamically weighting contributions based on modality reliability and input characteristics \cite{teng2024multi}.
Affective state estimation fuses audio, video, text, and physiological data using hybrid LSTM-Transformer architectures with task-specific class tokens, enabling simultaneous regression of arousal and valence through shared multimodal representations that capture temporal dependencies and cross-modal interactions \cite{marino2025temporal}.

\subsubsection{Structured Relational Methods}
\label{sec:plm_structured_relational}

Structured relational methods model interdependencies among psychological phenomena through graph-based architectures, knowledge graph embeddings, and relational reasoning frameworks.
These approaches capture complex relationships extending beyond individual instances to encompass network structures, temporal dynamics, and hierarchical dependencies in psychological data.

\paragraph{Single-Modal Text Methods}

\textbf{Graph Neural Network Architectures.}
Graph-based architectures integrate pre-trained language models with network structures to capture relational patterns in psychological text.

Personality detection benefits from psycholinguistic knowledge integration via tripartite graphs connecting posts, words, and LIWC (Linguistic Inquiry and Word Count) categories, where flow graph attention networks efficiently aggregate information across heterogeneous nodes using BERT-based initialization \cite{yang-etal-2021-psycholinguistic}.
The architecture uses three node types: user posts encoded by BERT, psychological lexicon entries, and LIWC dimensions, with graph attention mechanisms learning importance weights for edges connecting these heterogeneous nodes.
D-DGCN constructs dynamic graphs where posts are nodes with connections iteratively refined by a learn-to-connect module based on the output representations of the preceding graph convolutional layer, rather than relying on predefined edge weights \cite{yang2023orders}.

\textbf{Knowledge Graph Construction and Completion.}
Knowledge graphs provide a structured representation of psychological domain knowledge by organizing entities and their semantic relationships into relational triple formats.
Pre-trained language models serve as the computational backbone for constructing such graphs, enabling domain-specific knowledge graphs that encode relationships among mental health conditions and related entities from large-scale text corpora \cite{10.4108/eai.29-3-2024.2347315}.
Beyond graph construction, pre-trained models also support knowledge graph completion by treating entity and relation descriptions as textual sequences and computing triple plausibility scores through transformer-based encoding \cite{yao2019kg}.

\paragraph{Multimodal Fusion Methods}
Multimodal approaches integrate text with complementary modalities through graph-based fusion architectures.
Drawing analysis systems employ ClipCap for generating textual descriptions of psychological drawing elements and construct knowledge graphs where Relphormer, a Transformer-based reasoning method, performs link prediction to infer mental states from visual-textual patterns through similarity-based entity matching \cite{lyu2025painting}.

\subsubsection{Generative Interactive Methods}
\label{sec:plm_generative_interactive}

Generative interactive tasks produce contextually appropriate responses in conversational settings, encompassing dialogue systems, therapeutic chatbots, and task-oriented interactions.

\paragraph{Single-Modal Text Methods}

\textbf{Retrieval-Based Generation.}
Retrieval methods integrate external knowledge to enhance response quality in emotional support dialogues.
Early approaches retrieve domain knowledge from psychological literature through BERT-based semantic matching in knowledge graphs, enabling structured question answering over mental health concepts via template-based reasoning \cite{guo2021mental}.
Advanced methods explore both retrieval-based and generative strategies for knowledge integration, where the generative approach queries COMET for plausible knowledge triplets and incorporates the resulting commonsense relations into BART-based generation through soft positional encoding and masked self-attention \cite{shen2022knowledge}.
Recent generative retrieval approaches introduce Residual Identifiers (ResID) that encode emotional contexts and support strategies as hierarchical identifiers, enabling dynamic retrieval for joint context prediction and response generation \cite{yang2025green}.

\textbf{Neural Sequence Generation.}
Pre-trained generative models enable end-to-end response generation through fine-tuning.
EmpBot introduces sentiment understanding and empathy-reinforced auxiliary losses into T5 training to recognize and echo emotional states \cite{zaranis2021empbot}.
GPT-2 and DialoGPT adapt to psychotherapy dialogues through transfer learning on domain-specific corpora combining Reddit mental health communities and therapy transcripts \cite{das2022conversational}.
Mental health support systems construct event-based knowledge graphs (MHKG) by extracting eventualities from ASER and contextual relations from Reddit corpus, enriching DialoGPT input sequences with retrieved neighboring nodes to improve response generation \cite{tong2023improving}.
T5-based architectures leverage text-to-text frameworks with parameter-efficient fine-tuning using LoRA for mental health dialogue datasets \cite{alathamneh2024artificial}.

\textbf{Personalized Generation.}
Persona modeling enhances emotional support by tailoring responses to seeker characteristics.
CoBERT employs a single shared BERT encoder to encode the context, persona, and candidate response, and applies a multi-hop co-attention mechanism to separately compute interactive matching between the context-response pair and the persona-response pair, aggregating the resulting features for persona-based empathetic response selection \cite{zhong-etal-2020-towards}.
PAL first fine-tunes a generative model to infer seeker persona sentences from conversation history, then incorporates the inferred persona into response generation via cross-attention, with a strategy-based controllable decoding method further steering outputs toward personalized emotional support \cite{cheng2023pal}.

\paragraph{Multimodal Fusion Methods}
Multimodal fusion integrates textual, acoustic, and visual signals to enhance emotional understanding in support systems.
Affective interaction architectures employ BERT for semantic and emotion recognition from text while extracting Mel-spectrogram features from speech through CNNs, using gated attention mechanisms to dynamically fuse semantic embeddings with acoustic representations for emotionally expressive speech synthesis via VITS \cite{yuan2025multimodal}.

\subsection{LLM-based Methods}
\label{sec:llm_based}

LLM-based methods contain billions to trillions of parameters and leverage pre-existing world knowledge to perform psychological assessments with minimal task-specific training data.
This paradigm fundamentally differs from PLM approaches through emergent capabilities in zero-shot reasoning and in-context learning.

\subsubsection{Classification and Regression Methods}
\label{sec:llm_classification_regression}

LLMs address classification and regression tasks through prompt-based adaptation rather than extensive fine-tuning.
This approach reduces annotation requirements but introduces challenges in prompt design, output calibration, and bias control.

\paragraph{Single-Modal Text Methods}

\textbf{Zero-Shot Learning Mechanisms.}
Zero-shot learning performs psychological assessments without task-specific training examples by relying on pre-trained knowledge and instruction-following capabilities.
Models process natural language prompts that specify the assessment task and output format.
Classification applications span diverse constructs.
GPT-4o performs psychological text classification including reported speech identification and conversational repair detection through validated iterative prompting \cite{bunt2025validating}, while GPT-4o, Claude 3.5 Sonnet, and Gemini 1.5 Pro assess mental health urgency from social media posts using adapted triage frameworks \cite{settanni2025assessing}.
Clinical interview analysis employs GPT-4o for affective symptoms and Llama 3 for anhedonia detection \cite{teferra2025leveraging}.
Personality assessment adopts two formulations.
Discrete classification through GPT-3 requires knowledge injection strategies to improve accuracy \cite{ganesan2023systematic}.
GPT-3.5 and GPT-4 predict Big Five personality scores from Facebook status updates through zero-shot inference, achieving accuracy comparable to supervised machine learning approaches \cite{peters2024large}, while conversational analysis through GPT-4 infers traits from free-form interactions \cite{peters2024large1}.
Integration of BFI-10 frameworks into prompts enhances inference accuracy \cite{zhu2025investigating}.
Regression tasks predict continuous measurements through direct numerical estimation.
Flan-T5 extracts substance use disorder severity levels from clinical notes \cite{mahbub2025decoding}, GPT-4 analyzes interview transcripts to predict Social Phobia Inventory scores \cite{ohse2024gpt}, and Qwen 2.5-72b estimates MADRS item scores from clinical interviews \cite{kebe2025llamadrs}.

\textbf{In-Context Learning Strategies.}
In-context learning provides demonstration examples within prompts to adapt models to task-specific patterns without parameter updates.
Few-shot prompting employs Chain-of-Thought reasoning for mental health classification \cite{patil2025cognitive}.
The ExDoRA framework combines demonstration retrieval with explanation generation through hybrid ranking strategies \cite{priyadarshana2025exdora}, while cascade architectures integrate GPT-3.5-Turbo with linear networks for depression detection \cite{zheng2024cascade}.
Demonstration selection critically affects performance.
XAI4LLM employs knowledge-guided design with domain-specific feature grouping \cite{nazary2024xai4llm}, suicide risk assessment uses temperature control for consistency and few-shot examples for extreme categories \cite{thomas2025large}, and digital phenotyping transforms numerical behavioral features into natural language descriptions \cite{yuan2025leveraging}.
Regression applications calibrate numerical outputs through demonstrations.
Parkinson's disease assessment employs few-shot prompting with Gemma-2 and Llama-3.1 to predict neuropsychiatric scores from speech content \cite{castelli2025detecting}.

\textbf{Prompt Engineering Techniques.}
Systematic prompt design balances task specification, domain knowledge integration, output format control, and bias mitigation.
Task-specific architectures integrate clinical frameworks.
Psycho Analyst combines DSM-5 criteria with PHQ-8 tools through dual-task prompts \cite{tang2024advancing}, Chat-Diagnose employs Chain-of-Thought reasoning with tweet selectors for context management \cite{qin2025explainable}, and MAIMS implements two-step processes where models simulate scale completion then classify based on results \cite{li2024zero}.
Optimization strategies refine prompts through iterative development.
Empirical studies identify construct definitions and task descriptions as critical elements, automatic prompt engineering combines human design with algorithmic optimization \cite{anglin2025improving}, and PTSD diagnostics employ heterogeneous templates for different variable types \cite{tu2024automating}.
Psychologically-informed design incorporates domain knowledge.
PsyCoT transforms personality detection into multi-turn dialogues following questionnaire structures \cite{yang-etal-2023-psycot}, Diagnosis of Thought implements three-stage frameworks for cognitive distortion detection through subjectivity assessment, contrastive reasoning, and schema analysis \cite{chen2023empowering}, and ADOS-Copilot integrates clinical standards with statistical priors \cite{jiang2024copiloting}.
Context-aware strategies handle long documents and complex interactions.
FMAN employs semantic fusion of stimulus-response pairs through task minimization instructions \cite{liu2025llm}, BDI-biased summarization retrieves theme-related sentences, generates summaries, then classifies item scores \cite{aragon2024delving}, and MBTI detection uses soft-label datasets derived through iterative probabilistic estimation of personality polarity \cite{li2025can}.

\textbf{Instruction Tuning Approaches.}
Instruction tuning adapts LLMs through supervised fine-tuning on instruction-formatted datasets, enabling smaller models to match larger zero-shot models while maintaining generalization.
Mental-LLM applies instruction tuning to six classification tasks including stress prediction and suicide risk grading \cite{xu2024mental}.
Domain-adaptive tuning incorporates psychological knowledge.
Depression detection through fine-tuned GPT-3.5 Turbo and LLaMA2-7B on social media posts demonstrates substantial accuracy gains with optimized hyperparameters \cite{shah2025advancing}, and PHQ-9 labeling employs both few-shot prompting and instruction tuning with SFT and DPO \cite{shao2025systematic}.
Alignment techniques enhance models through preference learning.
Personality prediction from counseling dialogues uses role-playing prompts with questionnaire decomposition, with small models surpassing large models after alignment while requiring less input content \cite{yan2024predicting}.
Synthetic generation addresses data scarcity.
Chain-of-thought prompting with Llama 3.2-3B extracts structured summaries then generates controlled synthetic data through style transfer \cite{kang2024synthetic}.

\paragraph{Multimodal Fusion Methods}

Multimodal approaches integrate text, audio, and visual signals across communication channels.
Psychology-guided representation learning extracts theoretically-informed features.
Traits Run Deep employs psychology-guided prompts for semantic extraction, integrating text with audio and visual signals through text-centric fusion networks with chunk-wise projectors, cross-modal connectors, and ensemble regression heads \cite{li2025traits}.
Asynchronous pattern modeling addresses temporal misalignment.

\subsubsection{Structured Relational Methods}
\label{sec:llm_structured_relational}

LLM-based structured relational methods transform graph-based psychological modeling through natural language interfaces and emergent reasoning capabilities.
These approaches address network formation, knowledge graph construction, temporal dynamics modeling, and hierarchical relationship extraction without requiring specialized graph neural architectures.

\paragraph{Single-Modal Text Methods}

\textbf{Graph Structure Understanding through Linguistic Representation.}
LLMs process graph structures through natural language descriptions rather than specialized encoders.
Network formation simulation employs agent-based frameworks where LLM agents receive structured prompts describing local neighborhoods and demographic attributes to make connection decisions \cite{gkartziosmodeling}, while GPT-4o mini agents engage in social media environments through plan-execute-reflect cycles producing networks with homophily and reciprocity patterns \cite{schneider2025learning}.
Cognitive network analysis uses GPT-3.5 to generate associations and emotional valence ratings for STEM concepts, revealing lower clustering coefficients and sparser network structures compared to human networks \cite{haim2025cognitive}.
Personality analysis employs feature-activated dual-perspective networks where LLMs activate feature words through reverse generation combined with LIWC dictionaries to construct graphs capturing both semantic associations and structural dependencies \cite{mao2025feature}.

\textbf{Multi-hop Reasoning Mechanisms.}
Multi-hop reasoning chains connect evidence across inferential steps for complex assessment.
Network game simulation employs multi-step strategic reasoning where GPT-5-nano agents with MBTI personalities analyze historical behaviors, infer neighbor intentions, predict interaction outcomes, and make cooperation decisions in iterated prisoner's dilemma across network topologies \cite{qiu2025networkgames}.
Cognitive pathway extraction decomposes cognitive behavioral therapy frameworks where ERNIE 3.0 performs hierarchical classification identifying ABCD theory components and GPT-4 generates pathway summaries \cite{jiang2024ai}.

\textbf{Knowledge Enhancement Techniques.}
Knowledge enhancement integrates external psychological knowledge through structured representations and retrieval mechanisms.
Psychological counseling systems design ontology models using improved convolutional residual networks for relation triple extraction, with low-rank adaptation fine-tuning Qwen1.5-7B for Text2Cypher translation and multi-agent systems dividing labor across knowledge operations \cite{liu2024psychological}.
Prior elicitation employs GPT-5 to generate binary judgments on variable pair inclusion in psychological networks, translating responses into edge inclusion probabilities for Bayesian graphical modeling \cite{sekulovski2025llm}.

\textbf{End-to-End Knowledge Graph Construction.}
End-to-end construction automates graph building from raw data through LLM-driven extraction and reasoning.
Automated hypothesis generation combines LLM semantic understanding with causal graph reasoning where the LLMCG framework extracts causal knowledge from literature and generates novel hypotheses based on graph structure \cite{tong2024automating}.
Depression stigma knowledge graphs integrate AI-driven chatbot interviews with LLM causal relationship extraction, using ontology construction and entity resolution to form causal networks \cite{meng2025deconstructing}.
Large-scale mental health knowledge graphs employ LLM-powered extraction from biomedical literature, producing graphs with over 10 million relationships enriched with conditional statements, baseline characteristics, and confidence scores \cite{gao2025large}.

\textbf{Temporal Relationship Modeling.}
Temporal modeling captures psychological state evolution through sequential analysis and longitudinal integration.
EvolvTrip constructs perspective-aware temporal knowledge graphs that track characters' evolving beliefs, desires, and intentions throughout extended narratives, enabling Theory-of-Mind reasoning over long-horizon psychological trajectories \cite{yang2025evolvtrip}.
RELATE-Sim models long-term relationship dynamics by staging turning-point scenes between persona-aligned LLM agents, inferring interpretable commitment states and emotional appraisals to capture how dyadic psychological states transition over consequential life events \cite{yue2025relate}.
TASA addresses temporal knowledge evolution in educational psychology by maintaining structured student persona profiles and incorporating a continuous forgetting curve with knowledge tracing to dynamically update each learner's mastery state \cite{wu2025teaching}.

\paragraph{Multimodal Fusion Methods}

Multimodal approaches integrate text with complementary modalities for comprehensive psychological assessment.
Personality-guided depression detection employs LLMs to transform discrete personality features into contextual descriptions where P3HF applies Hypergraph-Former architectures modeling high-order cross-modal relationships with event-level domain disentanglement separating public patterns from private context-specific information \cite{fu2025personality}.
Drawing-based psychoanalysis employs multi-step MLLM reasoning decomposing drawings into hierarchical levels, using reinforcement learning-trained modules to extract object-specific features and integrate HTP knowledge base for psychological profiling \cite{ma2025reasoning}.
Intention prediction integrates offline knowledge graphs with online reasoning graphs from real-time perception, where LLMs perform reasoning over combined structures to interpret human intentions \cite{zhou2024enhancing}.

\subsubsection{Generative Interactive Methods}
\label{sec:llm_generative_interactive}

Generative interactive methods leverage sophisticated reasoning and multi-agent architectures to deliver psychological support. These approaches transform therapeutic dialogue through structured reasoning, self-verification, ensemble strategies, and generation constraints that ensure clinical safety and appropriateness.

\paragraph{Single-Modal Text Methods}

\textbf{Chain-of-Thought Reasoning.}
Chain-of-thought reasoning breaks down complex psychological processes into explicit intermediate steps, producing interpretable rationales aligned with clinical frameworks.
ESCoT models emotional support by identifying emotions, understanding their causes, and applying regulation strategies, enriching dialogue datasets with emotion chains and strategy rationales \cite{zhang-etal-2024-escot}.
For motivational interviewing alignment, models first predict the next MI strategy as internal reasoning before generating therapist responses that strictly follow that strategy \cite{sun2025rethinking}.
CFEG \cite{chen2024cause} introduces universal CoT prompt templates guiding stepwise emotion cause inference while integrating external knowledge from COMET \cite{bosselut-etal-2019-comet}.

\textbf{Complex Reasoning Decomposition.}
Complex reasoning decomposition tackles multi-faceted psychological tasks by partitioning intricate cognitive operations into specialized sub-processes.
MetaMind coordinates three agents: a Theory-of-Mind Agent generates mental state hypotheses, a Moral Agent refines them using cultural and ethical constraints, and a Response Agent produces validated empathetic responses \cite{zhang2025metamind}.
HealMe guides clients through cognitive reframing in three stages: separating situations from thoughts, brainstorming alternative perspectives, and integrating reframed thinking with empathetic encouragement \cite{xiao-etal-2024-healme}.
INT simulates expert narrative therapists across four therapeutic stages using hierarchical state planning with retrieval-augmented generation, while tracking progression through two-level Innovative Moment classification \cite{feng-etal-2025-reframe}.
MIND facilitates self-to-self caring through four specialized agents operating in iterative workflows, employing chain-of-thought prompting and cumulative memory to support cognitive evolution \cite{chen2025mind}.
SweetieChat delivers emotional support by training on counselor-guided psychological strategies for diverse open-domain scenarios \cite{ye-etal-2025-sweetiechat}.

\textbf{Ensemble Reasoning.}
Ensemble reasoning coordinates multiple models or reasoning paths to enhance response quality through complementary perspectives.
ChatPal trains compact models through teacher-student learning where large LLMs generate diverse emotional support dialogues via iterative expansion and Diverse Response Inpainting \cite{zheng-etal-2024-self}.
Polaris 2 deploys a constellation architecture where a primary agent drives empathetic conversations while specialist agents handle safety-critical tasks \cite{mukherjee2024polaris}.

AutoCBT implements a CBT-oriented framework where a Counsellor agent dynamically routes requests either to respond directly or consult Supervisors for advice refinement \cite{xu2025autocbt}.
MultiAgentESC achieves training-free emotional support through iterative dialogue analysis, strategy deliberation, and response generation addressing one-to-many strategy selection \cite{xu2025multiagentesc}.
MAGI automates MINI-based psychiatric interviews by coordinating four agents for navigation, adaptive questioning, response validation, and diagnosis generation with Psychometric Chain-of-Thought reasoning \cite{bi-etal-2025-magi}.
MENTALER combines chain-of-thought diagnosis, exemplar-based knowledge retrieval, and strategy-guided counseling text generation through multi-role collaboration \cite{gu2024mentaler}.
Dual dialogue systems support mental health providers via human-in-the-loop multi-agent architecture offering response proposals, conversation analysis, summarization, and resource recommendations \cite{kampman2024multi}.

\textbf{Advanced Generation Constraints.}
Advanced generation constraints guide outputs to reflect specific psychological characteristics while maintaining safety boundaries through sophisticated control mechanisms.
PsychAdapter enables continuous psychological trait control through lightweight transformer modifications independent of prompting \cite{vu2026psychadapter}.
Client101 simulates depression and anxiety clients using prompt-configured GPT-4 chatbots based on clinical vignettes for psychotherapy training \cite{lozoya2025leveraging}.
The cognitive reframing system controls seven linguistic attributes through retrieval-enhanced in-context learning, with field studies revealing preference for empathic and specific reframes over overly positive ones \cite{sharma2023cognitive}.
Controllable generation frameworks balance safety and helpfulness using self-generated data with numerical control tokens for dynamic attribute adjustment \cite{tuan2024towards}.

\paragraph{Multimodal Fusion Methods}
Multimodal approaches integrate text with complementary modalities for comprehensive psychological analysis and response generation.
ZS-CoDR detects cognitive distortions and generates reasoning in patient-doctor conversations through zero-shot multimodal learning with a hierarchical architecture that processes audio, video, and textual signals \cite{singh2024deciphering}.
Flexible thinking frameworks employ reinforcement learning to adaptively select contextual reasoning aspects including visual scene, emotion, situation, and response strategy, operating over text and video inputs for multimodal emotional support conversations \cite{wang2025flexible}.
Avatar-based benchmarks combine text, speech audio, and facial vision modalities to generate empathetic responses with synchronized facial expressions and vocal tones \cite{zhang2025towards}.
Trustworthiness reinforcement concatenates video, audio, and text features and leverages the attention mechanism of large language models to dynamically weight cross-modal contributions, enabling reliable emotional support delivery \cite{le2026reinforce}.

\section{Challenges}
\label{sec:challenges}

Despite significant advances in applying computational methods to psychological assessment, several fundamental challenges persist.
These challenges limit clinical utility and generalizability.
This section examines obstacles grounded in empirical evidence from deployed systems.
We highlight gaps between research prototypes and real-world requirements.

\subsection{Interpretability Requirements in Clinical Decision-Making}
\label{sec:interpretability}

Section~\ref{sec:background} discussed the paradigmatic tension between psychological explanation and computational prediction.
Current technical implementations fail to bridge this gap even when clinicians accept predictive models.
Deep learning opacity poses barriers that extend beyond epistemological differences.

Early predictive models in psychiatry, such as suicide risk assessment using machine learning~\cite{passos2016identifying}, achieved reasonable accuracy.
However, the black-box nature of algorithms obscured how clinical variables interacted to generate individual predictions.
This limitation restricts clinicians' ability to understand and trust the decision-making process.
Recent studies achieved improved prediction accuracy using structural MRI~\cite{hong2021identification} and dynamic functional connectivity~\cite{xu2022identification}.
However, the interpretability challenge remained.
These machine learning models could identify discriminative features but provided limited insight into the causal mechanisms.
These mechanisms link brain alterations to suicidal behavior.
Without transparent reasoning chains linking features to established theoretical constructs~\cite{cronbach1955construct}, practitioners cannot validate model outputs.
They cannot determine whether models capture genuine clinical phenomena or exploit spurious correlations.

Attention mechanisms emerged as a popular solution for visualizing model decisions in depression detection and anxiety screening.
However, highlighting specific words like ``I feel tired'' in clinical interviews fails to explain the underlying causes.
Fatigue may stem from physiological causes, situational stress, or represent a core DSM-5 symptom of major depressive disorder.
This symptom requires sustained duration and functional impairment for diagnosis.
Post-hoc explanation methods like saliency maps provide limited causal insight.
Meanwhile, constrained architectures sacrifice predictive performance.
The tension between model complexity and interpretability remains unresolved.
This challenge extends to structured relational tasks where models must explain not only entity identification but also why specific relations are predicted.
For instance, distinguishing (insomnia, symptom\_of, depression) from (insomnia, comorbid\_with, depression) requires transparent reasoning about temporal context and causal mechanisms.

\subsection{Label Uncertainty and Multi-Rater Disagreement}
\label{sec:label_uncertainty}

Psychological constructs involve subjective judgment.
This subjectivity leads to challenges in social media-based mental health detection.
Social media data often lacks formal clinical validation.
Labels are derived from user self-reports rather than professional diagnoses.
Additionally, data collection suffers from inherent biases.
These biases include unequal participation across age groups and demographics.
For example, Reddit users are predominantly aged 18-49~\cite{adarsh2023fair}.
These challenges stem from two sources: the absence of clinical validation and systematic bias in data representation.

Standard machine learning pipelines resolve disagreement through majority voting or expert adjudication.
This process imposes single consensus labels on uncertain cases~\cite{aroyo2015truth}.
This approach discards valuable information about divergent interpretations.
It also encodes the biases of dominant annotator groups into training data~\cite{davani2022dealing}.
Models trained on aggregated labels learn to predict a single interpretation that may not exist.
These models fail to capture the inherent ambiguity in subjective judgments~\cite{uma2021learning}.

The underlying cause of such disagreements often lies in annotators' lived experiences and backgrounds.
These factors systematically influence their judgments~\cite{diaz2022crowdworksheets}.
Interpretations of emotional expression severity or appropriateness differ across annotators.
These differences depend on their personal experiences with mental health and professional backgrounds.
When models are trained on labels reflecting specific annotator perspectives, they may encode these subjective viewpoints.
They fail to learn generalizable patterns~\cite{geva2019we}.
In structured relational tasks, this problem manifests as disagreement on relation types even when annotators agree on entities.
Extracting symptom-disorder associations may yield divergent interpretations where one annotator labels a relation as causal while another views it as correlational.

Recent work has begun modeling annotator disagreement explicitly.
However, the field lacks frameworks for distinguishing meaningful ambiguity from measurement noise.
This gap is especially problematic in subjective tasks.
In these tasks, individual perspectives naturally differ.
Forcing consensus in such cases may eliminate clinically relevant information about diagnostic uncertainty.

\subsection{Privacy-Preserving Learning for Sensitive Data}
\label{sec:privacy}

Psychological data's sensitive nature produces tensions between utility and privacy.
Recent applications use longitudinal passive sensing from smartphones and wearables~\cite{chikersal2021detecting, jacobson2022digital}.
These applications collect rich behavioral traces.
These traces include location, social interactions, physiological signals, and communication patterns.
These studies implement basic privacy protections such as data de-identification and access controls.
However, the high-dimensional nature of such behavioral data increases re-identification risks.
This nature requires more robust privacy-preserving techniques.

Federated learning provides a decentralized training approach.
This approach keeps data local at each institution.
Recent mental health applications show privacy-preserving collaborative model development across healthcare organizations~\cite{lee2023privacy}.
However, healthcare institutions operate heterogeneous IT infrastructures with varying security policies.
This heterogeneity makes coordinated training technically challenging~\cite{khalil2024exploring}.
Gradient-based federated learning is vulnerable to model inversion attacks.
These attacks can reconstruct individual training examples from shared model updates.
Differential privacy mechanisms add noise to protect individual records.
However, they create accuracy-privacy tradeoffs.
These tradeoffs are often unacceptable for clinical applications.
In these applications, false negatives carry serious consequences~\cite{yang2025deep}.

Psychological computing operates in regulatory ambiguity.
Existing frameworks do not address the unique challenges of mental health applications~\cite{iqbal2022regulatory}.
Inferred mental states are often more sensitive than the raw behavioral data from which they are derived.
Yet no consensus exists on quantifying or mitigating such risks.
Current regulations like HIPAA and GDPR primarily protect explicitly collected health data.
However, they offer limited guidance for algorithmically inferred psychological attributes.
The field lacks established protocols for determining appropriate accuracy-privacy tradeoffs in psychological assessment contexts.
Without clear ethical guidelines, developers and clinicians remain uncertain about accountability.
This uncertainty arises when AI systems make incorrect inferences about mental states~\cite{meadi2025exploring}.

\subsection{Cross-Cultural Validity and Bias}
\label{sec:cultural_validity}

Computational models exhibit cultural biases from Western over-representation in training data~\cite{kordzadeh2022algorithmic}.
Most publicly available mental health detection datasets derive from English-language social media or North American and European clinical populations.
Yet these models deploy globally without adequate cross-cultural validation.
The problem extends beyond language.
It includes variations in how constructs manifest across cultures~\cite{triandis1989self, kirmayer1989cultural}.

Cultural norms shape both learned behavioral signatures and ground truth labels.
Depression detection models may associate direct emotional disclosure with symptoms.
This association reflects Western communication norms.
Such models fail in cultures valuing emotional restraint.
In these cultures, distress manifests through somatic complaints or social participation changes.
Research comparing Chinese and American populations demonstrates that somatic symptom reporting varies significantly across cultural contexts and clinical status~\cite{yen2000cross}.
Current computational models lack mechanisms to account for such culturally specific symptom presentations.
Recent work has developed culturally adapted approaches using localized lexicons and domain-specific fine-tuning~\cite{han2020knowledge, ren2021sentiment}.
However, personality trait detection models trained predominantly on Western populations risk misclassifying normative collectivist behaviors as extreme profiles.

Multilingual models trained through translation or cross-lingual transfer capture linguistic diversity.
However, they do not capture deeper cultural differences in construct validity.
Fine-tuning on small target population samples provides limited benefit.
This limitation occurs when fundamental feature-construct relationships differ~\cite{kordzadeh2022algorithmic}.
The field lacks comprehensive frameworks for assessing whether computationally measured constructs maintain equivalent meaning across cultural groups.

Culturally diverse datasets with adequate sample sizes remain scarce.
Collecting such data requires multilingual capabilities, culturally informed annotation protocols, and validation against culture-appropriate clinical instruments.
Without addressing these validity concerns, computational psychological assessment risks exporting Western conceptualizations as universal standards.

\subsection{Long-term Personalization and Model Adaptation}
\label{sec:personalization}

Psychological states evolve dynamically over time.
Yet most computational models lack mechanisms for continuous individual-level adaptation after deployment.
Early work modeling substance use disorder relapse~\cite{witkiewitz2007modeling} highlighted the need to capture relapse as a dynamic process using longitudinal data.
However, existing approaches often rely on static population-level models.
Recent applications monitoring depression treatment response~\cite{chikersal2021detecting} demonstrate this challenge.
Models can be trained on 16-week longitudinal data.
However, they require frequent ground truth assessments that are impractical in real-world settings.
This requirement limits their ability to track individual trajectories over extended periods.
Foundation models for health outcome prediction~\cite{renc2024zero, placido2023deep} show promise in forecasting disease trajectories.
However, they do not incorporate mechanisms for personalized model updating as individual circumstances change.

Personalized models adapting to individual users face catastrophic forgetting~\cite{kirkpatrick2017overcoming}.
In this phenomenon, learning from new data degrades performance on previous patterns.
Models must adapt to genuine psychological state changes while maintaining core assessment capabilities.
Distinguishing signal from noise in limited person-specific data proves difficult.
This difficulty arises when individuals provide biased self-reports~\cite{stone2003patient}.
It also occurs when behavioral changes reflect situational rather than psychological factors.

Privacy considerations complicate personalization~\cite{xu2021federated}.
Modeling trajectories requires retaining historical data.
This retention increases privacy risks.
Online learning approaches update models incrementally without storing raw data.
However, these approaches struggle with temporal dependencies inherent in psychological processes~\cite{ijcai2018p369}.
Predicting cognitive decline requires modeling multi-year trends.
This requirement makes it difficult to balance adaptation speed with stability.
Structured relational tasks face additional complexity as relations evolve over time.
The relation (patient, exhibits, depression) valid at intake may no longer hold at discharge.
Current models lack mechanisms to track such temporal changes in relational structures.

The field lacks frameworks for determining when and how models should adapt~\cite{parisi2019continual}.
Methods are needed for several purposes.
First, they should detect when updates are warranted versus when changes reflect measurement noise.
Second, they should quantify uncertainty in personalized predictions based on limited individual data.
Third, they should balance personalization benefits against overfitting risks.
These risks involve fitting to patterns not reflecting meaningful psychological change.

\section{Future}
\label{sec:future}

While current LLM-based approaches have demonstrated substantial capabilities in psychological applications, several promising research directions remain underexplored.
This section outlines four key areas where future work can advance the field.

\subsection{Emerging Technical Frontiers}

Current multimodal models process different modalities through separate encoders before combining them at later stages, but this approach may miss the intrinsic correlations between linguistic expressions, facial dynamics, vocal prosody, and physiological signals.
Future models could learn \textbf{unified representation spaces} by training on naturally synchronized data, such as therapy sessions where speech, video, and physiological monitoring occur simultaneously.
The central challenge is developing pretraining objectives that align with psychological constructs rather than generic vision-language tasks.
Another promising direction involves integrating symbolic knowledge structures with neural components.
Pure neural approaches excel at pattern recognition but often fail at compositional reasoning about psychological states and their causal relationships.
\textbf{Graph neural networks} operating over psychological knowledge graphs could provide more interpretable inference and enable counterfactual reasoning that current end-to-end models cannot handle.
In these graphs, nodes represent constructs like emotions or symptoms while edges encode theoretical relationships between them.
Privacy considerations also demand attention.
Standard federated learning approaches may not suffice for psychological data, where small sample sizes and high sensitivity create unique challenges.
Differential privacy mechanisms need refinement to balance utility and privacy in clinical settings, while \textbf{on-device learning} could allow models to adapt to individuals without transmitting personal information, though this requires solving efficient personalization with limited local compute resources.

\subsection{Methodological Innovations}

Observational data dominates psychological applications, but correlation-based models cannot distinguish causal mechanisms from spurious associations.
Integrating \textbf{causal discovery algorithms} with LLMs could help identify plausible causal structures from longitudinal data.
Techniques like Granger causality and structural equation modeling, combined with learned representations, might reveal whether social isolation precedes depression or vice versa in specific populations.
The main obstacles include handling confounders, selection bias, and the high dimensionality of multimodal data.
\textbf{Active learning strategies} offer another opportunity to improve assessment efficiency.
Rather than using fixed question sets, systems could dynamically select the most informative questions based on previous responses.
An LLM might generate follow-up questions that disambiguate between competing diagnostic hypotheses, though this requires balancing exploration of diverse information with exploitation of high-confidence assessments.
This connects to the broader challenge of \textbf{continual learning} and lifelong personalization.
Mental states evolve over time, and models need to adapt to individual trajectories without catastrophic forgetting.
Meta-learning approaches could facilitate rapid adaptation by learning initialization points that require minimal fine-tuning, but a fundamental question remains: what aspects should be personalized versus what should remain shared across individuals?
Emotion expression patterns might benefit from personalization while general psychological knowledge should likely remain consistent.

\subsection{System and Ecosystem Development}

Moving from research prototypes to deployed systems requires addressing integration challenges at multiple levels.
Future systems should support effective \textbf{human-AI collaboration} rather than attempting full automation.
This means moving beyond simple decision support to mixed-initiative interaction where both human and AI contribute their strengths.
The technical work involves designing interfaces that make AI reasoning transparent and allow humans to efficiently provide context, while evaluation must consider not just accuracy but also trust, workload, and communication quality.
Such collaborative systems will function better within standardized ecosystems that enable \textbf{interoperability} across platforms.
The current proliferation of proprietary mental health tools creates fragmented environments where data and models cannot be easily shared.
Technical standards for psychological data formats, model interfaces, and evaluation protocols would help, though the challenge lies in accommodating diverse data types while ensuring backward compatibility as standards evolve.
Federated infrastructure could provide a foundation, but governance structures must also manage shared resources and balance standardization with innovation.
\textbf{Real-time closed-loop intervention systems} represent another important direction.
Current applications typically assess states and provide recommendations but do not systematically measure intervention effects and adjust strategies accordingly.
Closed-loop systems would continuously monitor outcomes and adapt interventions based on measured effects.
This requires low-latency inference, online learning algorithms that update safely without destabilizing behavior, and robust outcome measurement from noisy behavioral signals.
Control-theoretic frameworks from robotics could inform the design of stable feedback loops, though ethical safeguards against harmful adaptations remain critical.

\subsection{Emerging Application Frontiers}

The integration of LLMs with \textbf{immersive technologies} opens new therapeutic modalities.
Virtual and augmented reality combined with LLMs could create environments for exposure therapy where AI agents dynamically generate scenarios based on patient responses, or social skills training with virtual characters providing real-time feedback.
The technical work involves integrating language models with 3D environment generation, embodied agent control, and multimodal sensing while maintaining coherent long-term interactions.
The concept of \textbf{digital mental health twins} offers a framework for long-term personalized modeling.
These computational models would simulate an individual's psychological dynamics by integrating data from multiple sources, allowing clinicians to test intervention strategies computationally or enabling individuals to explore hypothetical scenarios.
Building accurate digital twins requires advances in personalized modeling, uncertainty quantification, and validation against longitudinal outcomes.
The long-term vision involves models that accumulate knowledge over a lifetime and provide continuity across different care providers.
Most current work focuses on individuals, but psychological phenomena also emerge at group and societal levels.
Future research could apply LLMs to understand \textbf{collective psychological dynamics} including emotions, social norms, and cultural patterns from large-scale data.
Agent-based models where LLM-powered agents interact could simulate social psychological phenomena like conformity or polarization.
The methodological challenge involves validating that computational models capture genuine social processes rather than data artifacts, while ethical considerations include avoiding surveillance applications and ensuring that insights benefit communities rather than enable manipulation.

\section{Conclusion}
\label{sec:conclusion}

The rapid expansion of AI-driven psychological research has created a fragmented landscape where similar computational techniques are independently developed across isolated domains.
This survey addresses this challenge by introducing the first systematic taxonomy that organizes psychological computing tasks based on computational characteristics rather than application domains.
Our framework identifies four fundamental task types: classification for discrete state identification, regression for continuous measurement, structured relational modeling for capturing interdependencies, and generative interactive systems for personalized engagement.
Through comprehensive analysis of over 300 representative works spanning the pre-trained model era and the large language model era, we trace the evolution from task-specific feature engineering to transfer learning and few-shot adaptation capabilities.
We provide systematic coverage of datasets, evaluation metrics, and benchmarks while examining fundamental challenges including interpretability, privacy preservation, cultural validity, and clinical utility that distinguish psychological computing from conventional AI applications.
This computational perspective reveals transferable methodological patterns previously obscured by domain-centric organization, enabling systematic knowledge transfer across subfields.
Future research should prioritize multimodal integration, personalization mechanisms balancing population and individual patterns, theory-aligned interpretability, cross-cultural validation, and dynamic adaptation systems for real-world deployment.
By organizing the field around computational commonalities rather than application boundaries, this survey establishes a foundation for accelerated progress in computational psychology and improved mental health outcomes at scale.

\bibliography{custom}

% \appendix

% \section{Example Appendix}
% \label{sec:appendix}

% This is an appendix.

\end{document}